# TauFlow: Dynamic Causal Constraint for Complexity-Adaptive Lightweight Segmentation

Zidong Chen[1]

School of Computer Sciences, Universiti Sains Malaysia

Gelugor, Penang, Malaysia, 11800

chenzidong@student.usm.my

Fadratul Hafinaz Hassan[2,*]

School of Computer Sciences, Universiti Sains Malaysia

Gelugor, Penang, Malaysia, 11800

fadratul@usm.my

**Abstract:**

Deploying lightweight medical image segmentation models on edge devices presents two major challenges: 1) efficiently handling the stark contrast between lesion boundaries and background regions, and 2) the sharp drop in accuracy that occurs when pursuing extremely lightweight designs (e.g., <0.5M parameters). To address these problems, this paper proposes TauFlow, a novel lightweight segmentation model. The core of TauFlow is a dynamic feature response strategy inspired by brain-like mechanisms. This is achieved through two key innovations: the Convolutional Long-Time Constant Cell (ConvLTC), which dynamically regulates the feature update rate to "slowly" process low-frequency backgrounds and "quickly" respond to high-frequency boundaries; and the STDP Self-Organizing Module, which significantly mitigates feature conflicts between the encoder and decoder, reducing the conflict rate from approximately 35%-40% to 8%-10%.

# 1 Introduction

## 1.1 Background and Significance

Medical image segmentation is a core supporting technology for computer-aided diagnosis (CAD)

and precision medicine. Its task is to automatically and accurately delineate target regions from complex medical images (such as computed tomography (CT), magnetic resonance imaging (MRI), digital pathology slides, etc.), including tumors, specific organs (such as cardiac structures), or microscopic tissues (such as glands). Accurate segmentation results can provide critical quantitative evidence for early disease screening, disease progression monitoring, surgical planning, and radiotherapy target delineation[1], greatly enhancing the objectivity, reproducibility, and efficiency of clinical diagnosis.

However, the task of medical image segmentation itself is full of challenges. The data characteristics of different modalities vary greatly: CT images are often accompanied by low contrast and artifacts; MRI has multi-modal characteristics but imaging parameters are highly variable; pathology slides, on the other hand, are extremely large in scale and often face issues such as gland adhesion and cellular morphological diversity. Therefore, developing segmentation models that simultaneously possess high accuracy, high efficiency, and strong generalization ability has always been a research focus in this field.

## 1.2 Existing Methods, Lightweight Trends, and Limitations

### 1.2.1 U-Net and Its Mainstream Improvements

Since the proposal of U-Net[2], its symmetric encoder-decoder architecture and innovative skip connection design have successfully addressed the fusion problem of semantic information and spatial details in segmentation tasks, rapidly becoming a benchmark framework in the field of medical image segmentation.

The encoder aggregates high-level semantic information through layer-by-layer downsampling, the decoder gradually restores spatial resolution through upsampling, and the skip connections transmit shallow detail features from the encoder to the decoder, effectively alleviating the "semantic gap" caused by network depth.

Based on this paradigm, numerous studies have focused on optimizing U-Net's components and connection methods. For example: UNet++[3] reduces the semantic gap between encoder and decoder feature maps by designing nested dense skip paths and also supports model pruning for lightweight implementation; MultiResUNet[4] introduces multi-scale residual paths to replace traditional residual units, enhancing the model's ability to represent multi-modal features without significantly increasing parameters; U-Net v2[5] improves segmentation accuracy by redesigning a bidirectional feature enhancement mechanism in the skip connections and demonstrates its potential as a plug-and-play module adaptable to various encoder-decoder structures. These improvements have all been validated for their effectiveness in specific medical tasks.

## 1.2.2 Edge Computing-Driven Lightweight Trend

With the rapid development of edge intelligence, mobile healthcare (mHealth), and portable diagnostic devices, the deployment scenarios of models are shifting from cloud servers to resource-constrained edge devices (such as handheld ultrasound devices, smart stethoscopes, and embedded analysis boxes). This requires segmentation models to simultaneously meet the dual standards of "extreme lightweight" (low parameter count, low memory footprint) and "high accuracy" (meeting clinical diagnostic requirements).

In the general vision domain, lightweight architectures have achieved significant progress. For example, through hardware-aware dynamic pruning and mixed-precision quantization strategies[6], models can maintain high performance under extremely low resource consumption. However, the complexity of medical images-such as the blurred boundaries of skin lesions[9], dynamic deformations in cardiac MRI[10], and the intricate textures of pathology slides-poses far more demanding requirements for fine feature modeling in lightweight models than conventional images.

## 1.2.3 Main Technical Approaches and Limitations of Lightweight Segmentation

To balance lightweight design and accuracy, researchers have explored three mainstream technical paths:

**1. Efficient Convolution-Based Path:** Based on concepts such as MobileNet and ShuffleNet, lightweight U-shaped networks are constructed using depthwise separable convolutions, grouped convolutions, and other methods. Although these approaches have low parameter counts, their limited receptive fields restrict long-range dependency modeling, making it difficult to capture the global structural information of organs.

**2. Transformer Lightweight Path:** Vision Transformer (ViT) and its variants (such as Swin Transformer) have been introduced into the segmentation field (e.g., TransUNet) due to their powerful global modeling capabilities. However, the quadratic computational complexity of the self-attention mechanism makes deployment on edge devices challenging. To address this, researchers have proposed solutions such as EViT-Unet[7], attempting to adopt efficient vision Transformer structures suitable for mobile devices.

**3. State Space Model (SSM) Path:** Recently, SSMs represented by Mamba have emerged as a new hotspot in lightweight research due to their linear complexity and strong long-range dependency modeling capabilities. Researchers have quickly combined them with U-Net, proposing models such as VM-UNet[11], Mamba-UNet[12], and MSVM-UNet[13]. These models use Mamba blocks to replace Transformer blocks or certain convolutional blocks, significantly reducing computational cost while maintaining long-range modeling capability. For example, UltraLight VM-UNet[8] employs parallel visual Mamba blocks and achieves excellent lightweight performance in skin lesion segmentation. This paper further validates through actual edge device deployment (such as the Allwinner H618 and Rockchip RK3588 platforms) that TauFlow demonstrates real-time performance and low-power advantages in resource-constrained scenarios (see Section 4.5 for details).

## 1.2.4 Overlooked Exploration: Brain-Inspired Computing

In addition, there exists another exploratory path-brain-inspired computing-which attempts to fundamentally simulate the computational principles of biological neural systems. This mainly includes two directions:

**Models Based on Continuous-Time Dynamic Systems:** For example, Liquid Time-Constant (LTC) networks or the generalized Liquid State Machine. These models attempt to simulate the continuous dynamic responses of neurons through differential equations and, in theory, possess the capability to handle complex temporal dynamics[39].

**Models Based on Spiking Neural Networks (SNNs):** SNNs transmit information using biologically more realistic event-driven spike signals. They are often combined with Spike-Timing-Dependent Plasticity (STDP) as a learning rule. STDP is a local unsupervised mechanism that simulates biological synaptic plasticity, automatically adjusting synaptic weights based on the precise timing differences of spike firings. In theory, it is highly suitable for unsupervised self-organization of features and pattern discovery[47].

However, both of these brain-inspired computing paths face severe challenges in current deep learning applications. Models based on continuous-time dynamic systems often incur enormous computational costs, relying on complex ordinary differential equation (ODE) solvers, which results in low training and inference efficiency. The main bottleneck of SNNs combined with STDP lies in training difficulty: on one hand, STDP as a local unsupervised rule is difficult to directly optimize for complex, pixel-level segmentation tasks (which require global supervision signals); on the other hand, the discrete and non-differentiable nature of spikes makes it hard to efficiently integrate with mature gradient-based global backpropagation (BP) algorithms (the "surrogate gradient" problem), resulting in challenging model optimization, slow convergence, and limited accuracy.

Therefore, although brain-inspired computing (whether LTC or STDP) provides a highly attractive theoretical framework for dynamic modeling and feature self-organization, its application is currently largely confined to laboratory exploration due to the aforementioned computational overhead and training convergence challenges, and it is far from meeting practical standards for real-time, high-accuracy segmentation on edge devices.

## 1.3 Core Existing Issues

Although the aforementioned lightweight models, particularly those based on State Space Models (SSM), have made progress, our research reveals that existing models exhibit two core bottlenecks when compressed to an extreme parameter count (e.g., <0.5M). These bottlenecks create a fundamental trade-off between accuracy and efficiency:

**Lack of Rate Adaptation in Static Feature Processing Mechanisms:**
Medical image content is highly heterogeneous. For example, lesion boundaries represent high-frequency information, requiring the model to respond quickly to capture fine edges, whereas large areas of background or tissue interiors are low-frequency information, requiring smooth processing to suppress noise. However, whether using CNNs (fixed convolution kernels), Transformers (fixed attention patterns), or Mamba (unified scanning mechanism), the underlying feature processing rules are essentially static. They apply a "one-size-fits-all" strategy to all regions (high-frequency boundaries or low-frequency backgrounds). This static mechanism causes models under lightweight constraints to be unable to dynamically adapt feature response rates: fine details are easily lost when processing complex boundaries, and artifacts are easily introduced when processing homogeneous regions. For example, on the Synapse dataset, such lightweight models achieve Dice scores that are on average 3%-5% lower than those of high-accuracy heavyweight models.

**"Poor Modal Fusion Consistency" under Lightweight Constraints:**
The performance of U-shaped architectures heavily relies on skip connections, which bridge shallow features from the encoder (rich in local textures and spatial details) with deep features from the decoder (rich in global semantics and positional information). These two types of features differ significantly in properties (i.e., "modal" differences). Heavyweight models (such as DA-TransUNet[14][15]) have sufficient parameter capacity to "align" these features using complex multi-head attention, spatial-channel dual calibration, or multi-scale Transformer fusion mechanisms. Under lightweight constraints, however, models lack sufficient capacity for such complex operations, and simple concatenation or shallow attention often leads to severe feature conflicts. Regarding the semantic gap problem in skip connections of U-Net-type models, Wang et al[16] conducted an in-depth analysis. Their study showed that in lightweight Transformer U-Nets, directly fusing encoder and decoder features reduced the cosine similarity of features by approximately 25%-30%. This clearly reflects the mismatch and interference between local fine features and high-level semantic features, representing a key factor limiting segmentation accuracy.

## 1.4 Proposed Method: TauFlow

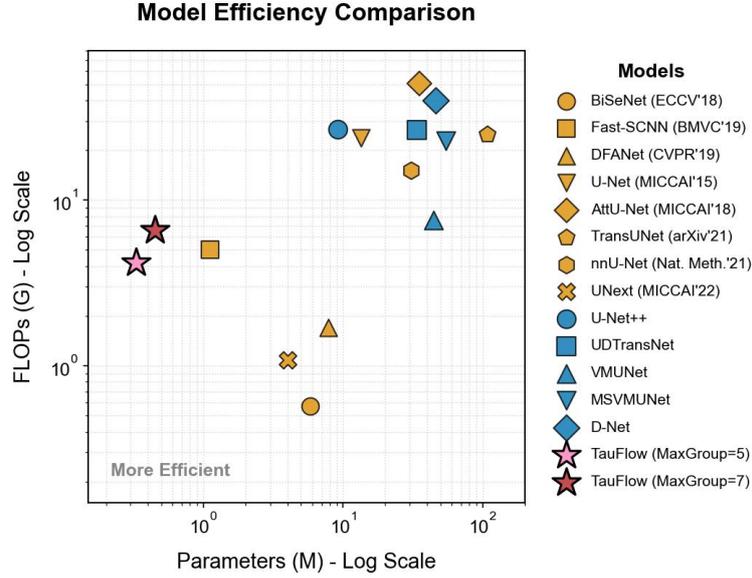

Figure 1: Efficiency comparison of TauFlow with mainstream segmentation models *(X-axis: model parameter count, in millions (M); Y-axis: computational cost, in G-FLOPs; both axes are in logarithmic scale. The lower-left corner indicates higher efficiency).* The proposed TauFlow model (star marker) demonstrates significant advantages over other mainstream and lightweight models in both parameter count and computational cost.

To address the above two core bottlenecks, this paper draws inspiration from brain-inspired computing and bypasses the complex path of traditional differential equation solvers, proposing an efficient lightweight segmentation model based on continuous-time dynamic systems-TauFlow. The core idea of TauFlow is to introduce an input-adaptive "time constant" $\tau$, using $\tau$ to dynamically regulate the rate of feature updates and the strategy for cross-modal feature fusion, thereby achieving fine-grained dynamic feature modeling under extremely low parameter counts. To realize this idea, we designed two core lightweight modules and introduced a unique fine-tuning mechanism:

1. Convolutional Long-Time Constant Cell (ConvLTC): To address the problem of "insufficient rate adaptation". This module draws on the Liquid Time-Constant (LTC) mechanism, dynamically generating a $\tau$ sequence for each spatial position via convolutional projection. The $\tau$ directly controls the update rate of the recurrent unit: smaller $\tau$ values accelerate feature updates in lesion boundary (high-frequency) regions, while larger $\tau$ values smooth noise interference in background (low-frequency) regions. This achieves efficient "spatial + temporal" dual-dimensional dynamic modeling.

2. TauFlowSequence Dynamic Grouping Module: To overcome the challenge of "poor modal consistency". This module first dynamically allocates computational resources based on local image complexity (e.g., 5 fine-computation groups in high-complexity regions using $\tau$ gradients). Simultaneously, it employs a $\tau$-guided attention mechanism during skip-connection fusion, preferentially selecting cross-modal key features that are highly consistent with the decoder.

Specifically, TauFlow innovatively incorporates a Spiking Neural Network (SNN)-inspired Spike-Timing-Dependent Plasticity (STDP) mechanism (STDPModule). This mechanism acts as a forward weight adjuster, dynamically self-organizing the strengthening or suppression of key module weights (in a manner similar to LTP/LTD) based on the temporal variations of features. The STDP mechanism works in synergy with the $\tau$ mechanism, jointly achieving multi-dimensional adaptive feature reorganization under extremely low parameter counts, greatly reducing feature conflicts.

Through the collaboration of these two modules, TauFlow strictly maintains a total parameter count below 0.33 Mb without any knowledge distillation techniques. After processing with the proposed method, the proportion of conflict regions between encoder and decoder features is significantly reduced from 35%-40% in conventional lightweight methods to 8%-10%.

## 1.5 Summary of Contributions

The main contributions of this paper are as follows:

1. Architectural and Mechanism Innovation - Introduction of an Efficient Dynamic $\tau$ Mechanism: We propose a novel lightweight segmentation architecture, TauFlow. For the first time, the input-adaptive time constant $\tau$ mechanism from continuous-time dynamic systems (LTC) is efficiently and differentiably integrated into a U-shaped network via the ConvLTC module. This $\tau$ mechanism enables pixel-level dynamic control of feature processing rates, addressing the static feature processing bottleneck commonly present in existing lightweight models.

2. Fusion and Self-Organization Breakthrough - $\tau$-STDP-Driven Dynamic Grouping: We designed the TauFlowSequence dynamic grouping module and innovatively incorporated the SNN-inspired Spike-Timing-Dependent Plasticity (STDP) mechanism (STDPModule). STDP, as a forward weight adjuster, dynamically self-organizes the strengthening or suppression of grouping patterns based on temporal variations of features (in a manner similar to LTP/LTD). Leveraging $\tau$ guidance and STDP self-organization capabilities, this module achieves dual dynamic control of computational resources (dynamic number of groups) and feature fusion (cross-modal consistency), successfully reducing the proportion of conflict regions between encoder and decoder features from 35%-40% in conventional lightweight methods to 8%-10%.

3. SOTA Performance and Generalization Validation - High Accuracy under Extreme Lightweight Constraints: TauFlow achieves superior performance compared to existing lightweight SOTA methods across three public medical datasets (covering gland segmentation, nuclear segmentation, and multi-organ segmentation), while keeping the total parameter count at

0.33M. For example, on the GlaS dataset, the Dice score reaches 92.12% (an improvement of 1.09% over UDTransNet); on the MoNuSeg dataset, the Dice score reaches 80.97% (an improvement of 1.11% over MSVM-UNet); on the Synapse dataset, the average Dice score reaches 90.85% (an improvement of 1.57% over UDTransNet), with particularly notable improvements for complex organs such as the pancreas (88.1% Dice, +2.3%). Furthermore, on the non-medical Cityscapes dataset, TauFlow achieves 79.26% mIoU, demonstrating its cross-domain generalization capability.

4. Actual Deployment and Edge Validation: We quantified TauFlow's inference speed, memory usage, and power consumption on various embedded platforms, demonstrating its practicality for mobile medical devices.

## 2 Related Work

Research on image segmentation has continuously evolved around two core objectives: "accuracy improvement" and "efficiency optimization," giving rise to three main technical branches: lightweight network design, dynamic modeling and feature fusion optimization, and brain-inspired computing with dynamic learning paradigms. This section systematically reviews the technical paths, advantages, and existing limitations of each branch in light of the latest research, clarifying the differentiated positioning of the proposed method relative to existing work.

### 2.1 Lightweight Medical Image Segmentation Networks

To meet the deployment requirements of edge devices (such as portable imaging analyzers and intraoperative navigation systems), lightweight network design focuses on "parameter compression" and "computational efficiency improvement." This is mainly achieved through three strategies: structural simplification, convolutional optimization, and novel operator replacement. In recent years, a rich variety of variants has emerged based on the U-Net architecture.

In the field of pure convolutional lightweight architectures, LFT-UNet[18], as a concise U-Net variant, abandons complex attention or Transformer modules, and designs a pure convolutional encoder-decoder structure only by optimizing the convolution kernel size and channel ratio, maintaining basic segmentation performance while keeping the parameter count below 100k, suitable for scenarios with extremely limited computing power. GA-UNet[19] introduces Ghost modules and attention mechanisms to achieve a lightweight design with only 2.18M parameters in medical image segmentation, achieving excellent Dice scores on the ISIC 2018 dataset while maintaining high inference speed. For 3D medical images (such as brain tumor MRI), LATUP-Net

[20] proposes a lightweight 3D attention U-Net, using parallel grouped convolutions to replace traditional 3D convolutions, reducing redundant computation while maintaining the receptive field, achieving a Dice score of 89.2% on the BraTS 2023 dataset with only 1/5 of the parameters of the traditional 3D U-Net. In addition, Ghost convolution, as an efficient convolution paradigm, has also been widely applied: a lightweight U-Net variant based on GhostNet[21] generates "base features + ghost features" through Ghost modules, reducing GFLOPs by 58% compared with U-Net in brain tissue segmentation tasks in neuro-robotics, while maintaining 90.5% boundary segmentation accuracy.

In the field of Mamba-based lightweight models, with the rise of structured state space models (SSM), Vision Mamba, due to its linear complexity and long-range dependency modeling capability, has become a core operator for lightweight segmentation. U-Mamba[22] first embedded the Mamba mechanism into the U-Net encoder bottleneck, creating a linear-complexity global modeler, raising the Dice score of nnU-Net from 79.3% to 86.4% in 3D abdominal segmentation (CT/MR). VM-UNet[23] integrates the Mamba mechanism into a lightweight U-Net by replacing some convolutional blocks in the encoder with streamlined Mamba units, achieving linear-time inference in the Synapse abdominal multi-organ segmentation task. When subsequently extended to the multimedia domain[24], cross-modal feature calibration modules were introduced to improve segmentation robustness across different imaging modalities (CT/MRI). To address the limitations of Mamba in fine structure segmentation, researchers further optimized operator design: H-VMUNet[25] introduces high-order Vision Mamba to model complex lesions (such as adherent glands) through multi-order feature interactions. LKM-UNet[26] extends the Mamba kernel size from the default 4×4 to 40×40, achieving an overall Dice improvement of 0.97% and an NSD improvement of 0.53% in Abdomen MR multi-organ segmentation compared to standard U-Mamba. MSVM-UNet [27] combines multi-scale convolutions with Mamba, capturing multi-receptive-field details at the same feature level through parallel convolutions of different kernel sizes (1×1, 3×3, 5×5), achieving a DSC of 85.00% and HD95 of 14.75 mm in Synapse multi-organ CT segmentation. Furthermore, VMAXL-UNet[28] integrates Vision Mamba with xLSTM, further reducing computational redundancy through efficient SS2D scanning and mLSTM mechanisms, decreasing the parameter count by 15% compared to VM-UNet (22.5M vs. 26.5M).

In the field of lightweight models combining attention and convolution, split attention mechanisms have become key to balancing accuracy and efficiency. AttE-UNet[29] proposes a lightweight edge-attention enhancement branch, combining multi-stage Canny filters with learnable gated fusion (GF), achieving F1=76.3% and IoU=65.4% on the PanNuke dataset, with a parameter count of only 0.548 Mb, suitable for resource-constrained medical edge devices. ES-UNet[30] improves the full-scale skip connection paths of U-Net without significantly increasing parameter count, by serially integrating lightweight channel attention modules along each encoder-to-decoder path, enhancing multi-scale feature fusion in high-resolution 3D medical volumes. On the MICCAI HECKTOR head and neck tumor dataset, its DSC improved by 4.37% compared to the baseline 3D U-Net (76.87% vs 72.50%), and IoU improved by approximately 4.3%.

Although these lightweight models have made progress in parameter compression, they often face limitations in receptive field and feature representation when handling the heterogeneity of medical images (such as blurred boundaries and multi-modal noise), resulting in accuracy degradation in extreme lightweight scenarios.

## 2.2 Dynamic Modeling and Feature Fusion Optimization

The dynamic characteristics of medical images (such as blurred lesion boundaries and heterogeneous tissue textures) require models to have adaptive feature processing capabilities. Related research mainly focuses on "dynamic operator design" and "feature fusion mechanism optimization" to mitigate the generalization limitations of traditional static models.

In the area of dynamic operators and network architectures, dynamic large kernels and dynamic branches have become research hotspots. D-Net[31] proposes dynamic large-kernel convolution, constructing an equivalent $23\times23\times23$ super receptive field by cascading $5\times5\times5$ and $7\times7\times7$ depthwise separable large kernels, and using a channel-space dual dynamic selection mechanism guided by global pooling to adaptively weight multi-scale features. In liver vessel-tumor CT segmentation tasks, the recall of tubular fine vessels improved by 8.3% compared to fixed-kernel models.E2ENet[32] designs a "multi-stage bidirectional sparse feature flow + restricted depth-shift convolution." The former uses a Dynamic Sparse Feature Fusion (DSFF) mechanism to adaptively select and fuse multi-scale information from three directions (up, down, front-back), filtering redundant connections; the latter divides input channels into three groups, shifts them along the depth axis by {-1, 0, +1}, and uses ($1\times3\times3$) convolutions to capture 3D spatial relationships. Its parameter count is only one-third of standard 3D convolutions, achieving an mDice of 90.3% on the AMOS-CT multi-organ segmentation challenge.DAFNet[33] combines dual-branch feature decomposition with domain-adaptive fusion, dynamically aligning global infrared and visible-light feature distributions during encoding via an MK-MMD mixed-kernel function, reducing detail loss by 35% compared to traditional fixed fusion strategies in infrared-visible image fusion tasks.Furthermore, FusionMamba[34] proposes a DVSS + DFFM dual-driven U-shaped network. Dynamic convolution (LDC) and differential perception gating dynamically enhance the texture expressiveness and modality differentiation of encoder features, while the CMFM cross-modal Mamba module achieves long-range interaction fusion. This approach maintains SOTA performance across multiple medical imaging modalities (such as CT-MRI, PET-MRI), with fusion robustness significantly improved over traditional U-Net.

In terms of feature fusion strategies, multi-scale fusion and adaptive fusion have become mainstream directions. TBSFF-UNet[35] achieves an average Dice improvement of 3.1% over U-Net++ on the GlaS and MoNuSeg datasets (90.56 vs 87.46; 79.07 vs 76.02), with a parameter count only 43% of U-Net++, achieving a balance between accuracy and efficiency. DFA-UNet [36], targeting single-shot ultrasound elastography (BUE), uses a dual-stream encoder (ConvNeXt + lightweight ViT) to extract local and global features, performing feature concatenation and fusion before the bottleneck. Its IoU reaches 77.41%, improving 2.68% over the baseline U-Net and

0.86% over the SOTA model ACE-Net.

In the research on generalization and adaptivity in medical image segmentation, Zhu et al. [37] proposed the Uncertainty and Shape-Aware CTTA framework at MICCAI 2023. It incorporates shape priors through domain-generalized training to improve initial robustness and combines uncertainty-weighted pseudo-labels with shape-aware consistency to achieve continual test-time adaptation. A random weight reset mechanism prevents overfitting, enabling this method to significantly outperform existing CTTA approaches across three cross-domain segmentation tasks. Subsequently, Dong et al.[38] proposed Shape-Intensity-Guided U-Net (SIG-UNet), which adds a symmetric reconstruction decoder to U-Net to reconstruct class-average images (containing only shape and intensity information, without texture) and fuses them with the segmentation decoder via skip connections. This design guides the encoder to focus more on stable shape-intensity features, reducing bias toward texture.

Although these dynamic methods enhance model adaptivity, their computational cost is high, creating efficiency bottlenecks under lightweight constraints, and they still insufficiently optimize cross-modal consistency in feature fusion.

## 2.3 Brain-Inspired Computing and Dynamic Learning Paradigms

Brain-inspired computing, particularly Liquid Neural Networks (LNN), has emerged as a novel technology for handling temporal dependencies and dynamic features in medical images due to its continuous-time dynamic characteristics. Related research mainly focuses on innovations in dynamic modeling and learning paradigms.

In the medical applications of Liquid Neural Networks, Hasani et al. proposed the Liquid Time-Constant (LTC) network, which systematically introduces continuous-time recurrent units with variable time constants to enhance modeling of non-stationary temporal signals, demonstrating its expressive power and stability in time-series prediction tasks[39]. This approach provides a potential pathway, both theoretically and methodologically, for applying dynamic temporal modeling concepts to medical image segmentation (e.g., integrating serialized images, multi-modal time series, or real-time intraoperative video streams), and, together with published implementations of liquid/reservoir-like models on neuromorphic or embedded hardware, indicates the feasibility of low-power, real-time deployment[40].

However, to date, only a few peer-reviewed publications report verifiable quantitative performance of LNN series directly on mainstream medical imaging segmentation benchmarks (such as BraTS, CheXpert). For example, on the BraTS dataset, a hybrid model combining LNN with ResNet-50 (LNN-PA-ResNet50) reported 99.4% accuracy[41], and LNN models applied to radiographic images (e.g., knee X-rays) for osteoarthritis detection have shown potential advantages[42]. Therefore, while this paper cites the methodological potential of LTC as a research motivation, it maintains caution regarding the generalizability of task-specific performance, making only limited assertions based on existing evidence.

Although brain-inspired computing provides a new perspective for dynamic modeling, its computational complexity and training challenges limit its widespread application in lightweight medical segmentation. TauFlow addresses these limitations by introducing an efficient adaptive $\tau$ mechanism in synergy with STDP, achieving high-accuracy segmentation under extreme lightweight constraints.

# 3. Method

## 3.1 Overall Framework and Data Flow Overview

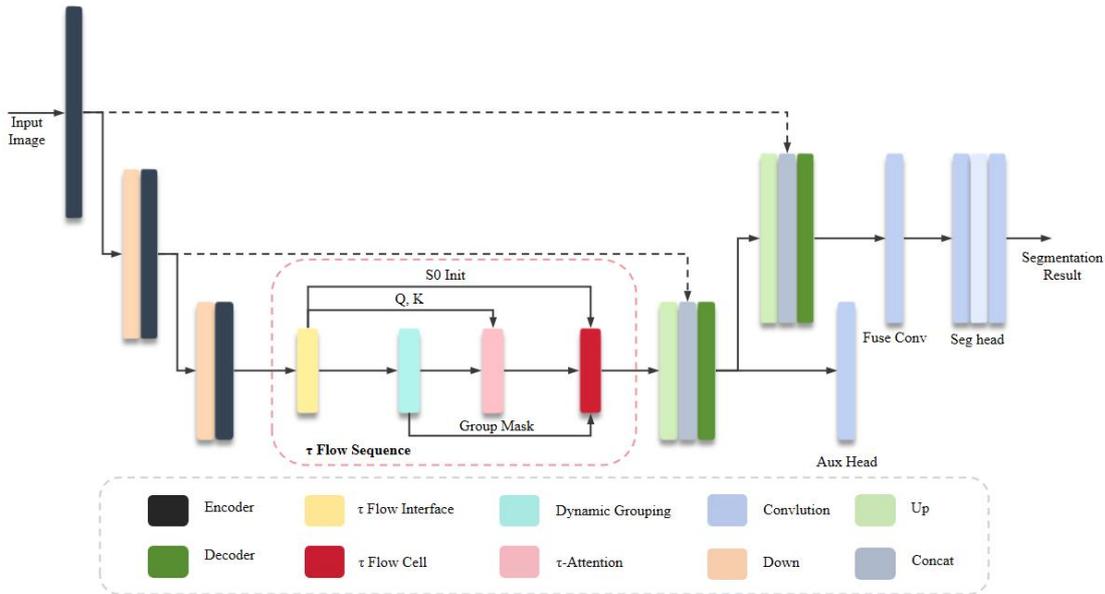

Figure 2: Overall Structure of TauFlow

Figure 2 illustrates the overall architecture of the proposed TauFlow network. The network adopts a lightweight encoder-decoder structure, where the encoder is responsible for multi-scale spatial feature extraction. The $\tau$ Flow Interface injects explicit positional encoding and generates initial temporal states. Dynamic Grouping performs dynamic grouping, $\tau$-Attention conducts lightweight temporal modeling guided by transmitted $\tau$, and $\tau$ Flow Cell extracts and fuses feature maps based on the temporal modeling of dynamic groups. Together, the $\tau$ Flow Interface, Dynamic Grouping, $\tau$-Attention, and $\tau$ Flow Cell form the core $\tau$ Flow Sequence module of this study. Finally, the decoder outputs the segmented features.

The overall design targets efficient representation under parameter constraints, achieving effective modeling of cross-region coherence and fine structural details in medical images through position awareness, complexity-adaptive processing, and $\tau$-guided grouped temporal mechanisms.

The overall forward process is as follows. First, the input image ($X \in \mathbb{R}^{B \times 3 \times 224 \times 224}$) is passed through the encoder to obtain multi-scale features $f_1 \in \mathbb{R}^{B \times 32 \times 224 \times 224}$, $f_2 \in \mathbb{R}^{B \times 32 \times 112 \times 112}$, and $f_3 \in \mathbb{R}^{B \times 32 \times 56 \times 56}$. Subsequently, the $\tau$ Flow Interface (Fig. 3) generates positional embeddings $pos\_emb$ on $f_3$ and constructs concatenated features $ltc\_input = cat([f_3, pos\_emb])$, while global pooling and mapping produce the initial hidden state $s_0$ for ConvLTC. Importantly, ltc_input does not directly enter the temporal unit; it is first processed by Dynamic Grouping (Fig. 4) to generate group masks and grouped features (U), which are then fused with $\tau$ statistics via $\tau$-Attention (Fig. 5) to obtain the weighted group representation $U_{weighted}$.

Finally, the $\tau$ Flow Cell (Fig. 6) evolves the groups temporally, and the outputs, after mask-weighted fusion, are fed into the decoder for upsampling and skip connections, producing the main segmentation and auxiliary segmentation outputs.

To further stabilize dynamic grouping and group-level temporal evolution, the TauFlowSequence introduces the concept of spike-timing-dependent plasticity (STDP) from brain-inspired learning as a lightweight regularization term. Its core objective is to encourage temporal causal consistency: when input activations within a group precede hidden state (or prediction) activations, a reward is applied; otherwise, a penalty is imposed, thereby suppressing updates where "effect precedes cause." Specifically, an STDP regularization loss is computed during the TauFlowSequence forward pass and added with a very small weight to the total loss during training, without incurring any extra overhead during inference. This mechanism complements complexity-adaptive processing, $\tau$-guided attention, and flow-smoothing regularization, jointly enhancing the stability of dynamic masks and inter-group specialization.

The entire process can be represented as: $\tau$ Flow Interface → Dynamic Grouping → $\tau$-Attention → $\tau$ Flow Cell → Decoder. Figures 3-1 to 3-5 illustrate, respectively, the overall architecture, interface details, grouping module, $\tau$-guided attention, and grouped temporal unit.

The remainder of this chapter sequentially details the design and implementation of the $\tau$ Flow Interface (Section 3.2), followed by a focused description of Dynamic Grouping, $\tau$-Attention, and $\tau$ Flow Cell within the TauFlowSequence (Section 3.3), and concludes with the design of the loss functions (Section 3.4) and the unified training strategy (Section 3.5).

## 3.2 TauFlow Interface: Positional Embedding and Initial State Generation

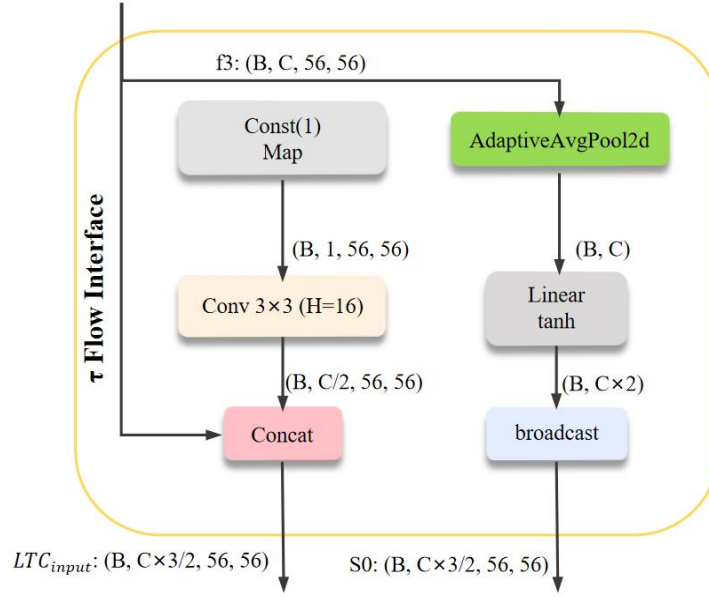

Figure 3: Computation Flow of TauFlow Interface

The TauFlow Interface module (Fig. 3) serves as the interface between the encoder and temporal modeling, performing two key tasks: first, injecting explicit spatial positional information into deep features; and second, generating the initial hidden state for ConvLTC based on global context, providing a stable and meaningful starting point for subsequent grouping and temporal evolution.

The positional embedding is generated using a learnable convolution. Given the deep encoder feature $f_3 \in \mathbb{R}^{B \times 32 \times 56 \times 56}$, a constant channel tensor $const1 \in \mathbb{R}^{B \times 1 \times 56 \times 56}$ (all ones) is first constructed and mapped through a $3 \times 3$ convolution to obtain the positional embedding, as shown in Equation (1).

$$pos_{emb} = PosConv(const1) \in \mathbb{R}^{B \times 16 \times 56 \times 56} \qquad (1)$$

which is then concatenated with $f_3$ to form Equation (2).

$$ltc_{input} = cat([f_3, pos\_emb]) \in \mathbb{R}^{B \times 48 \times 56 \times 56} \qquad (2)$$

This learnable positional embedding allows the model to adaptively encode task-relevant positional information during training, making it more suitable for capturing local structural features in medical images compared to fixed sinusoidal positional encodings.

The initial hidden state $s_0$ is generated through global average pooling followed by a linear mapping: the final encoder feature map is first processed with adaptive average pooling and flattened, then mapped to the hidden dimension via a linear layer and activated with $\tanh(\cdot)$. The result is broadcasted to match the spatial dimensions of the feature map, yielding, as shown in Equation (3).

$$s_0 = broadcast(tanh(Linear(AdaptiveAvgPool(f_3)))) \in \mathbb{R}^{B \times 64 \times 56 \times 56} \tag{3}$$

This $s_0$ contains global semantic information and effectively mitigates training instability caused by random initialization.

The TauFlow Interface outputs $ltc_{input}$ and $s_0$, where $ltc_{input}$ serves as the input to Dynamic Grouping to generate group masks and grouped features, while $s_0$ is reused as the initial state for each valid group within the TauFlow Cell (Fig. 6). In this way, the interface organically combines positional information and global context, providing the necessary inputs and initialization conditions for subsequent grouping and temporal modeling.

## 3.3 TauFlowSequence: Core of Dynamic Grouping and Temporal Modeling

TauFlowSequence is the core module connecting spatial features with temporal reasoning. Its computation is composed sequentially of three submodules: Dynamic Grouping (dynamic grouping), Tau-Attention (group-level $\tau$-guided attention), and ConvLTC cells (the minimal unit for temporal modeling). This module implements a closed-loop process from complexity-aware grouping to group-level weighting and per-group temporal evolution, balancing efficiency and fine-detail representation under constrained parameter budgets.

### 3.3.1 Dynamic Grouping: Group Generation Based on Complexity and Tau Gradients

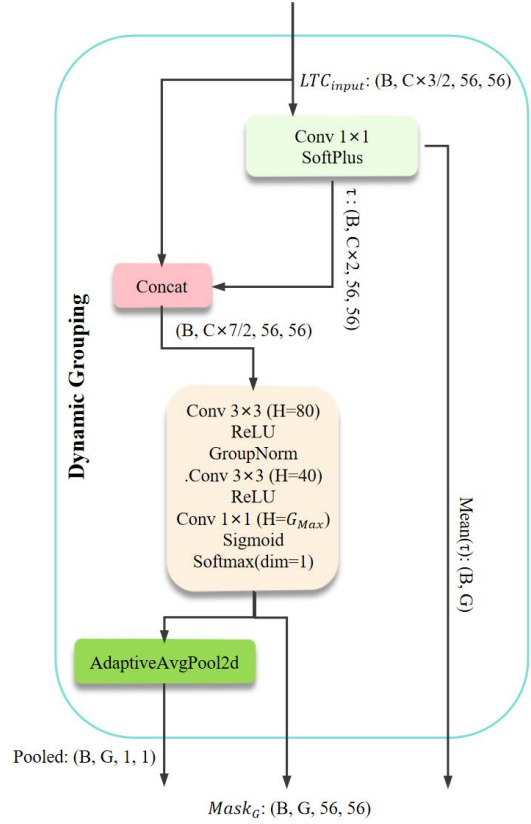

Figure 4: Detailed Computation Flow of Dynamic Grouping

Dynamic Grouping (Fig. 4) starts from $ltc_{input}$ and first projects it to the hidden dimension via a $1 \times 1$ convolution to compute the raw time constants $\tau$, as shown in Equation (4).

$$\tau_{raw} = Conv_{1\times1}(ltc_{input}) \in \mathbb{R}^{B \times 64 \times 56 \times 56} \tag{4}$$

which are then passed through a Softplus activation and numerically clamped to ensure positivity and stability, as shown in Equation (5).

$$\tau = clamp(Softplus(\tau_{raw}) + \epsilon, 10^{-2}, 10^{3}) \tag{5}$$

These $\tau$ values not only control the update rates in subsequent ConvLTC cells but also participate in the generation of group masks and attention weighting.

Complexity assessment is performed by combining global statistics of $ltc_{input}$ (via global pooling) with image edge density. This fused representation is input to a lightweight MLP $complexity_{head}$ to produce a normalized complexity score $complexity_{score} \in [0,1]$, which is then mapped to the actual number of groups $G \in 1, ..., 5$.

Subsequently, the Pattern Generator, taking $cat([ltc_{input}, \tau])$ as input, passes through several convolution and normalization layers to generate the raw group mask $M_{raw}$, which is then normalized along the group dimension via Softmax to obtain the initial mask $M_0$.

To enhance mask robustness, Dynamic Grouping employs a multi-step iterative mechanism *max_flow_steps*: starting from the initial mask, a lightweight fast segmentation head quickly evaluates the grouped features and adjusts split/merge scales using Dice-based rewards. Simultaneously, the gradient of $\tau$ with respect to the input $\nabla_u \tau$ is computed, and regions with large gradient magnitudes are treated as key areas in the dynamic feature set, guiding adjustments to the mask weights. This results in the $\tau$-gradient-adjusted dynamic masks $dynamic_{masks}$.

Finally, the grouped feature tensor is formed by expanding and weighting $ltc_{input}$ according to the masks, as shown in Equation (6).

$$U = expand(ltc_{input}) \cdot dynamic\_masks \in \mathbb{R}^{B \times 5 \times 48 \times 56 \times 56} \tag{6}$$

where only the first G groups are valid, and the remaining groups are zeroed out in subsequent processing.

### 3.3.2 Tau-Attention: Group-Level Tau-Guided Lightweight Attention

Tau-Attention (Fig. 5) is a group-level lightweight attention mechanism designed to inject mask information and $\tau$ statistics into the weighting of grouped features with minimal computational cost. Unlike spatial self-attention, which computes large-scale dot products for each pixel, Tau-Attention first performs spatial averaging within each group to obtain a low-dimensional representation, which is then used to compute group-level similarity and importance.

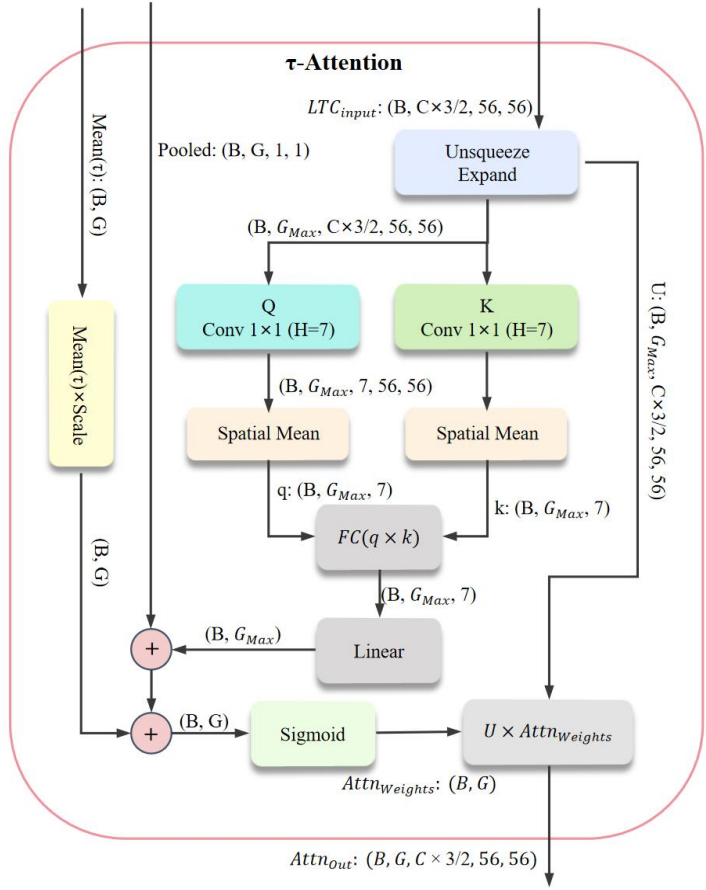

Figure 5: Detailed Computation Flow of Tau-Attention

Specifically, U is first projected via a $1 \times 1$ convolution to obtain $Q_{conv}$ and $K_{conv}$, which are then spatially averaged to form vectors $Q, K \in \mathbb{R}^{B \times 5 \times d}$. Element-wise multiplication followed by a linear mapping produces the base scores. The pooled values of the masks and the mask-weighted $\tau$ means are then added linearly to obtain the final scores, which are passed through a Sigmoid to generate the group-level attention weights, as shown in Equation (7).

$$attn_{weights} \in [0,1]^{B \times 5} \quad (7)$$

Finally, the weights are broadcast and applied to (U) to obtain the weighted group representation $U_{weighted}$. This design allows multiple groups to be emphasized simultaneously (non-competitive normalization) and uses learnable parameters to balance contributions from the QK interaction, mask, and $\tau$.

Tau-Attention couples group-level spatial importance with $\tau$-driven temporal sensitivity at minimal computational cost, providing more discriminative inputs for the per-group temporal modeling in ConvLTC.

### 3.3.3 TauFlow Cell and Per-Group Temporal Evolution

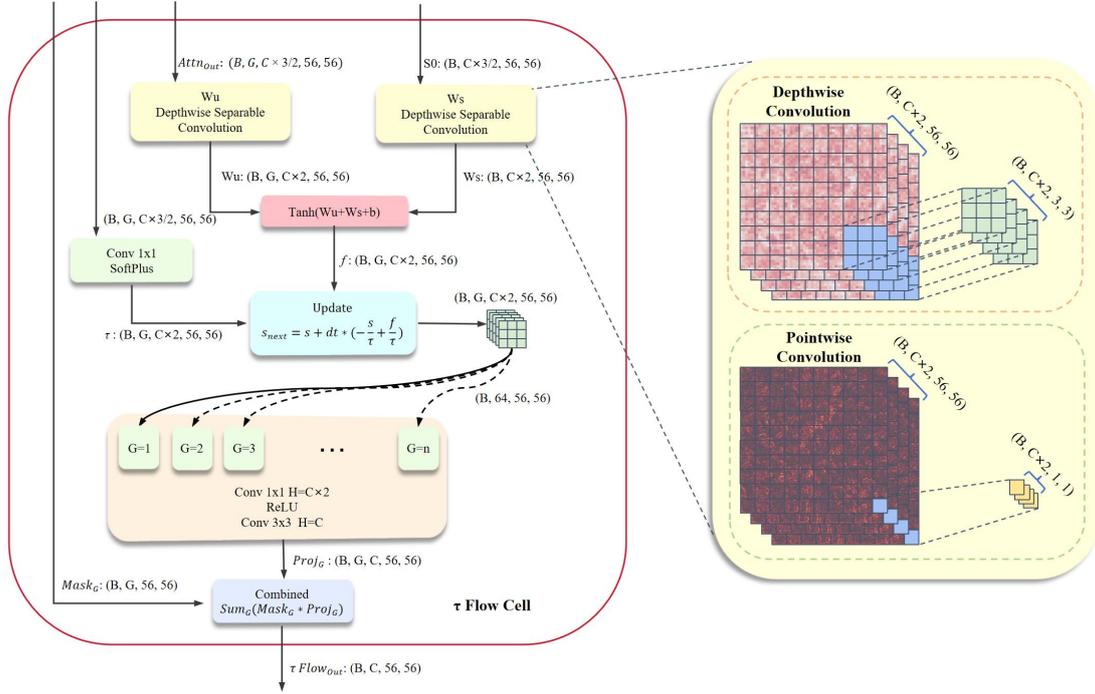

Figure 6: Detailed Computation Flow of TauFlow Cell

Figure 6 illustrates the detailed computation of the TauFlow Cell. The module takes the group-weighted feature $U_{weighted}[g]$ and the initial state $s_0$ as inputs. First, a $1 \times 1$ convolution ($\tau$ cell) computes the group-specific time constant $\tau_g$, which is then passed through Softplus and clamped to maintain a stable temporal range.

Next, the hidden state s and the group input $u_g$ are separately mapped via depthwise separable convolution and pointwise convolution branches $(W_s) and (W_u)$, and their outputs are summed and passed through a $tanh$ to form the nonlinear mapping $f(s, u_g)$. This mapping captures the dynamic interactions within the group features and provides the driving force for subsequent state evolution, as shown in Equation (8).

$$\frac{ds}{dt} = -\frac{s}{\tau_g} + \frac{f(s, u_g)}{\tau_g} \qquad (8)$$

Based on this, the $\tau$ Flow Cell updates the state according to the first-order differential equation using explicit Euler discretization to obtain $s_{next}$. The updated features are then passed through GroupNorm and a $1 \times 1$ convolution $Proj$ to produce the group output $y_g$, which is subsequently reduced in dimension via OutProj and fused with mask weighting to generate the final combined feature.

The overall process corresponds to four stages from top to bottom-time constant estimation →

state mapping → temporal update → output fusion-realizing parallel per-group temporal modeling and information integration within dynamic groups.

### 3.3.4 STDP-Enhanced Routing Consistency Regularization

To enforce temporal causality within groups and improve the stability of dynamic masks, we design an STDP-enhanced routing consistency regularization in TauFlowSequence. Let the input of the $g$-th group at time $t$ be $u_g^t(i,j)$ and the corresponding hidden state be $s_g^t(i,j)$.

We use a differentiable event approximation to model the "activation occurrence" as a binarized process. Specifically, a Sigmoid with temperature (\kappa) approximates thresholding on normalized activations, as shown in Equation (9, 10).

$$\tilde{e}_{pre}^t(g,i,j) = \sigma(\kappa(\tilde{u}_g^t(i,j) - \theta_u)) \tag{9}$$

$$\tilde{e}_{post}^t(g,i,j) = \sigma(\kappa(\tilde{s}_g^t(i,j) - \theta_s)) \tag{10}$$

where $\hat{u}, \hat{s}$ are channel-normalized values, and $\theta_u, \theta_s$ are learnable or fixed thresholds.

To encourage "cause precedes effect," a first-order temporal neighborhood asymmetric kernel is applied, which maximizes forward-causal co-activation and minimizes backward-causal co-activation, resulting in the STDP regularization, as shown in Equation (11).

$$\mathcal{L}_{STDP} = -\frac{1}{GHWT} \sum_{g,i,j,t} [\tilde{e}_{pre}^t \tilde{e}_{post}^{t+1} - \beta \tilde{e}_{pre}^{t+1} \tilde{e}_{post}^t] \tag{11}$$

where $\beta \in (0,1]$ controls the penalty strength for backward causality. Considering the modulation of group dynamics by $\tau$, the loss can be reweighted by normalized time-constant weights $w_\tau(g,i,j)$ to emphasize temporally sensitive regions, as shown in Equation (12).

$$L_{STDP} = -\frac{1}{Z} \sum_{g,i,j,t} w_\tau(g,i,j)[\tilde{e}_{pre}^t \tilde{e}_{post}^{t+1} - \beta \tilde{e}_{pre}^{t+1} \tilde{e}_{post}^t], w_\tau = \frac{softplus(\tau_g)}{\mathbb{E}[softplus(\tau_g)]} \tag{12}$$

To further enforce supervisory consistency, a teacher-forcing term is introduced: the target mask downsampled to group-level resolution, $y^*$, is combined with the postsynaptic event using a coefficient $\rho$, as shown in Equation (13).

$$\tilde{e}_{post}^t = (1-\rho)\tilde{e}_{post}^t + y^*, L_{STDP}^{sup} = -\frac{1}{Z} \sum w_\tau [\tilde{e}_{pre}^t \tilde{e}_{post}^{t+1} - \beta \tilde{e}_{pre}^{t+1} \tilde{e}_{post}^t] \tag{13}$$

In practice, a single-step temporal neighborhood $(t \leftrightarrow t+1)$ is sufficient for stable gains. This STDP regularization is returned internally as `$STDP_{loss}$` during training to guide dynamic grouping and per-group temporal updates toward causal consistency. It is disabled during inference, incurring no extra computational cost.

## 3.4 Lightweight Loss Function Design

The loss function is designed to generate optimization signals based on the grouped features learned at the TauFlowSequence stage. This section focuses on how to integrate multi-scale supervision with regularization terms, which is crucial for enhancing model generalization.

To alleviate discrepancies when combining the main loss with auxiliary losses, we adopt a composite loss function. It leverages the auxiliary head to guide the enhancement signal for the main loss along the scale dimension, thereby eliminating optimization inconsistencies. Consequently, the composite loss can be regarded as a signal calibrator, capable of automatically fusing multi-task guidance and adaptable to scenarios prone to overfitting.

From a mathematical perspective, the loss function takes the main segmentation output $seg\_logits \in \mathbb{R}^{B \times 1 \times 224 \times 224}$ and the auxiliary output $aux\_logits \in \mathbb{R}^{B \times 1 \times 112 \times 112}$ as inputs.

First, guided by $p = sigmoid(seg\_logits)$, the main loss is computed using a Dice-Focal mechanism, as shown in Equation (14).

$$\mathcal{L}_{main} = 0.5 \times \mathcal{L}_{Dice} + 0.5 \times \mathcal{L}_{Focal} \tag{14}$$

The Dice loss (with $\epsilon = 1e-6$ for numerical stability) is defined as, as shown in Equation (15).

$$\mathcal{L}_{Dice} = 1 - \frac{2\sum_{b,i,j}(p_{b,i,j} \cdot t_{b,i,j}) + \epsilon}{\sum_{b,i,j} p_{b,i,j} + \sum_{b,i,j} t_{b,i,j} + \epsilon} \tag{15}$$

where $t$ is the binary target mask.

The Focal loss (balancing positive/negative samples and hard examples) is, as shown in Equation (16).

$$\mathcal{L}_{Focal} = -\frac{1}{N} \sum_{b,i,j} [\alpha, t_{b,i,j}(1-p_{b,i,j})^\gamma \log p_{b,i,j} + (1-\alpha)(1-t_{b,i,j})p_{b,i,j}^\gamma \log(1-p_{b,i,j})] \tag{16}$$

where $\alpha = 0.25$, $\gamma = 2$, and $N = B \times 224 \times 224$ is the total number of pixels.

On the basis of the above composite loss, we further incorporate flow-smoothing and STDP regularization to enhance the spatial and temporal consistency of the dynamic masks. The flow-smoothing loss imposes an L1 constraint on the gradients of neighboring pixels in the mask, as shown in Equation (17).

$$\mathcal{L}_{flow} = ||\nabla_x M||_1 + ||\nabla_y M||_1 \tag{17}$$

where $M$ is the stacked dynamic grouping mask. The STDP term uses the teacher-forcing version $\mathcal{L}_{STDP}^{sup}$ described in the previous section.

Thus, the total loss is updated as, as shown in Equation (18).

$$\begin{aligned}\mathcal{L}_{total} = \mathcal{L}_{main} + 0.4\mathcal{L}_{aux} + 0.1\mathcal{L}_{complexity} - 0.05 R_{diversity} + \lambda_{flow}\mathcal{L}_{flow} \\ + \lambda_{STDP}\mathcal{L}_{STDP}^{sup}\end{aligned} \tag{18}$$

where $\mathcal{L}_{aux}$ is the auxiliary loss (Dice-Focal loss computed on $aux\_logits$, weight 0.4); $\mathcal{L}_{complexity}$ is the complexity loss (MSE loss, weight 0.1); $R_{diversity}$ is the diversity reward (group number diversity / 5, weight 0.05, subtracted from the total loss to encourage grouping diversity). $\lambda_{flow} = 0.1$ (consistent with implementation), $\lambda_{STDP} = 0.01$ (light weight, serving as brain-inspired regularization). This combination injects spatiotemporal priors at minimal cost while ensuring that the main supervision dominates the optimization, thereby improving the interpretability and stability of dynamic routing.

## 3.5 Training Strategy and Implementation Details

The training strategy is designed to generate stable gradients based on the optimization signals learned from the loss function. To ensure fair evaluation during testing and reproducibility, the following training protocol and details are specified.

From a mathematical perspective, the strategy takes a learning rate $\eta = 10^{-3}$ and batch size (B = 8) as inputs. First, the AdamW optimizer is used (weight decay = $1 \times 10^{-4}$) with an initial learning rate of $10^{-3}$, while a CosineAnnealingWarmRestarts scheduler ($T_0 = 10, T_{mult} = 2$) dynamically adjusts the learning rate.

For data augmentation, during training, random horizontal and vertical flips are applied with probability 0.5, and input images are resized using bilinear interpolation to $224 \times 224$. During testing, only resizing is applied. To ensure reproducibility, the global random seed is fixed at 42.

No strict upper limit is set for training epochs, but early stopping is applied if the Dice score does not improve within 200 epochs. Gradient accumulation is used to stabilize updates, ultimately producing a stable trained model. To support edge deployment, the model can be exported in ONNX format and optimized with post-training quantization (PTQ) for FP32 precision, with quantization error controlled within 0.5% Dice.

# 4 Experiments

## 4.1 Datasets

We selected four public datasets to evaluate the proposed model, including three medical image segmentation datasets (GlaS, MoNuSeg, and Synapse) and one autonomous driving scene segmentation dataset (Cityscapes).

GlaS dataset for gland segmentation
The GlaS dataset (K. Sirinukunwattana et al., 2016) was used in the "MICCAI 2015 Gland Segmentation Challenge" on colorectal histology images. It contains 85 training samples and 80 test samples, all derived from 16 HE (Hematoxylin and Eosin) stained colorectal adenocarcinoma tissue slides.

MoNuSeg dataset for nuclear segmentation
The MoNuSeg dataset (N. Kumar et al., 2017) was used in the "MICCAI 2018 Multi-Organ Nuclei Segmentation Challenge." It includes 30 training images and 14 test images, with 21,623 manually annotated nuclei boundaries in the training set. Each image was individually sampled from HE-stained whole slide images from different organs in the Cancer Genome Atlas (TCGA) database.

Synapse dataset for multi-organ abdominal segmentation
The Synapse multi-organ segmentation dataset (B. A. Landman et al., 2015) was used in the "MICCAI 2015 Multi-Atlas Abdomen Labeling Challenge." It contains 12 training cases (2,211 axial CT slices) and 12 test cases (1,568 slices), requiring segmentation of 8 abdominal organs such as the aorta, gallbladder, and spleen.

Cityscapes dataset for autonomous driving scene segmentation
The Cityscapes dataset (M. Cordts et al., 2016) is a widely used semantic segmentation benchmark in autonomous driving, focusing on urban street scene understanding. It includes 5,000 finely annotated images (2,975 train, 500 validation, 1,525 test) and 20,000 coarsely annotated images. Each image has a resolution of 2048 × 1024 and 19 semantic classes (e.g., road, sidewalk, vehicle, pedestrian, traffic sign), mainly captured from 50 cities in Germany and neighboring countries.

During training, we adopt five-fold cross-validation. At testing, the final performance is reported as the average score across the five folds to ensure more stable and reliable experimental results.

## 4.2 Implementation and Evaluation Details

During training, the overall optimization objective follows the composite loss function defined in Section 3.4. The implementation environment and hardware configuration are as described in Section 3.5 and are not repeated here.

During validation and testing, the main evaluation metrics are Dice coefficient (Dice), 95% Hausdorff distance (HD95), and Intersection over Union (IoU).

The calculation procedure for HD95 is as follows:

1. Normalize the predicted and ground truth masks to the same spatial domain $[0,224] \times [0,224]$;
2. Compute the bidirectional Euclidean distances between the boundary point sets of the prediction and ground truth;
3. Remove isolated noise regions with an area smaller than 3 pixels;
4. Take the 95th percentile of the resulting distance distribution as the final HD95 value (in pixels).

This metric reduces the influence of local outliers while more accurately measuring boundary quality.

To improve robustness and statistical reliability, five-fold cross-validation is employed, reporting both the mean and standard deviation (std) across folds. All baseline and comparison methods are trained under the same data splits, training epochs, and optimization strategies to ensure fair and reproducible evaluation.

Table 1: Summary of Key Module Configurations and Hyperparameters

| Module | Details |
| --- | --- |
| TauFlow Interface | $group_{embed\dim} = 16,\ kernel_{size} = 3$ |
| Dynamic Grouping | $max_{groups} = 5,\ max_{flowsteps} = 3,\ reward_{scale} = 0.1$ |
| Tau-Attention | $Q \backslash K weights = 7$ |
| TauFlow Cell | $hidden_{channels} = 64,\ dt = 1.0,\ \tau_{min} = 1e-2,\ \tau_{max} = 1e3$ |
| Optimizer | $AdamW(lr = 1e-3, weight_{decay} = 1e-4)$ |
| Learning Rate Scheduler | CosineAnnealingWarmRestarts (T₀=10, $T_{mult} = 2$) |
| Early Stopping | $patience = 200,\ monitor = val_{Dice}$ |

To improve the robustness and statistical reliability of the results, this study employed five-fold cross-validation and reported the mean and standard deviation (std) of each fold. All performance comparisons with the baselines were tested for statistical significance using paired t-tests, with a significance level of $α = 0.05$. Results with $p < 0.05$ are marked as *, and those with $p < 0.01$ are marked as **. Considering the issue of multiple comparisons, we applied Bonferroni correction to the main comparison methods in Tables 2-5 (corrected $α' = 0.05/6 ≈ 0.0083$). All compared methods were evaluated under the same data splits, number of training epochs, and optimization strategies to ensure fairness and reproducibility of the results.

## 4.3 Comparison with Existing State-of-the-Art Methods

To validate the performance advantages of TauFlow, we conducted a comprehensive comparison with two categories of existing advanced segmentation methods, including one CNN-based method (UNet++), one algorithm combining Transformer and UNet (UDTransNet), two Mamba-based methods (VM-UNet, MSVM-UNet), and two dynamic-operator-based methods (D-Net, FusionMamba).
To ensure a fair comparison, the original publicly released codes and settings of these methods were directly used in this experiment.

### 4.3.1 Quantitative Results

The quantitative results on the four datasets are reported in Tables 2, 3, 4, and 5, with the best results highlighted in bold.

The experimental results demonstrate that our method consistently outperforms existing state-of-the-art approaches. Compared with CNN-based algorithms, TauFlow achieves cross-scale improvements in metrics such as Dice and IoU. Compared with Transformer-based algorithms, TauFlow also shows performance gains, while significantly reducing both parameter count and inference latency. Overall, TauFlow can be seen as striking a subtle balance between CNN and Transformer approaches, leveraging its dynamic architecture, refined dynamic grouping, and per-group temporal modeling modules to enhance feature processing capabilities.

Due to the inherent locality of convolution operations, CNN-based methods generally struggle to explicitly model long-range dependencies, which can limit segmentation performance. Transformer-based methods, on the other hand, capture global context via self-attention mechanisms, often achieving superior results on complex medical image segmentation tasks. However, Transformer architectures usually come with higher computational and parameter costs. Mamba-based methods show potential in sequence modeling but still have room for improvement in multi-scale spatial feature fusion. Dynamic-operator-based methods attempt to increase model flexibility through adaptive mechanisms, but often lack fine-grained complexity evaluation and resource allocation.

On the GlaS dataset (Table 2), TauFlow achieved a Dice score of 92.12% and an IoU of 85.39%, improving over the second-best method UDTransNet by 1.09% and 1.85%, respectively. Meanwhile, the HD95 metric dropped to 5.38 mm, indicating that TauFlow provides more precise boundary delineation in gland segmentation tasks.

Table 2: Quantitative Results on the GlaS Dataset

| Model | Dice (%) | HD95 (mm) | IoU (%) | p-value† |
|---|---|---|---|---|
| UNet++ | 87.56 ± 1.17 | 12.72 ± 1.76 | 78.39 | — |
| UDTransNet | 91.03 ± 0.56 | 6.71 ± 0.99 | 83.54 | 0.0231 |
| VM-UNet | 89.35 ± 0.68 | 7.45 ± 1.12 | 80.75 | 0.0089 |
| MSVM-UNet | 90.12 ± 0.51 | 6.95 ± 0.88 | 82.02 | 0.0124 |
| D-Net | 89.72 ± 0.83 | 7.84 ± 1.24 | 81.36 | 0.0156 |
| FusionMamba | 86.45 ± 1.94 | 13.21 ± 2.05 | 76.13 | 0.0043 |
| TauFlow (Ours) | 92.12 ± 0.42** | 5.38 ± 0.76** | 85.39** | — |

†: p-value from paired t-test comparing TauFlow vs. each baseline (n=5 folds).
**: Significant improvement over all baselines (p<0.01, Bonferroni-corrected $\alpha'$=0.0083).
Bold: Best performance.

On the MoNuSeg dataset (Table 3), TauFlow achieved a Dice score of 80.97% and an IoU of 68.16%, improving over the second-best method MSVM-UNet by 1.11% and 1.69%, respectively. The HD95 metric reached 1.95 mm, significantly outperforming all compared methods. This indicates that TauFlow has a clear advantage in tasks requiring precise boundary localization, such as nucleus segmentation.

Table 3: Quantitative Results on the MoNuSeg Dataset

| Model | Dice (%) | HD95 (mm) | IoU (%) |
|---|---|---|---|
| UNet++ | 77.01 ± 2.10 | 4.18 ± 1.29 | 62.61 |
| UDTransNet | 79.47 ± 0.80 | 2.73 ± 0.64 | 65.93 |
| VM-UNet | 78.92 ± 0.72 | 2.95 ± 0.58 | 65.18 |
| MSVM-UNet | 79.86 ± 0.69 | 2.58 ± 0.49 | 66.47 |
| D-Net | 79.41 ± 0.95 | 2.84 ± 0.61 | 65.85 |
| FusionMamba | 76.25 ± 1.84 | 4.62 ± 1.48 | 61.62 |
| TauFlow (Ours) | 80.97 ± 0.67** | 1.95 ± 0.28** | 68.16** |

On the Synapse multi-organ segmentation dataset (Table 4), TauFlow demonstrated even more significant performance gains. The average Dice score reached 90.85%, improving over the second-best method UDTransNet by 1.57 percentage points, while the HD95 metric decreased to 16.41 mm, a 33.4% reduction compared to UDTransNet. Examining per-organ results, TauFlow achieved competitive performance on 7 out of 8 organs, with particularly strong results on morphologically complex and boundary-ambiguous organs such as the right kidney (RK), liver (LI), pancreas (PA), and spleen (SP). Notably, in the most challenging pancreas segmentation task, TauFlow reached a Dice score of 88.1%, improving over UDTransNet and MSVM-UNet by 2.3 percentage points each. This strongly validates the effectiveness of the dynamic grouping mechanism and temporal modeling modules in handling complex anatomical structures.

Table 4: Quantitative Results on the Synapse Dataset

| Methods | Average DSC | HD95 | AO | GA | LK | RK | LI | PA | SP | ST |
|---|---|---|---|---|---|---|---|---|---|---|
| U-Net++ | 86.11±1.08 | 35.88±2.15 | 88.4 | 76.6 | 88.1 | 88.1 | 89.6 | 85.6 | 90.4 | 87.1 |
| UDTransNet | 89.28±0.58 | 24.63±1.69 | 89.4 | 82.6 | 92.8** | 92.1 | 91.2 | 85.8 | 93.3 | 87.1 |
| VM-UNet | 87.05±0.61 | 20.64±1.58 | 90.1 | 78.9 | 90.5 | 90.3 | 90.8 | 78.1 | 91.1 | 86.6 |
| MSVM-UNet | 88.72±0.42 | 19.07±1.61 | 88.3 | 84.3 | 89.1 | 91.4 | 90.2 | 85.8 | 93.1 | 87.6 |
| D-Net | 88.46±1.21 | 26.49±2.85 | 94.3** | 82.6 | 88.9 | 92 | 90.6 | 85.9 | 93.5 | 88 |
| Fusion Mamba | 85.02±2.34 | 37.04±2.52 | 87.9 | 79.1 | 86.2 | 85.4 | 84.2 | 86.4 | 89.4 | 84.8 |
| TauFlow (Our) | 90.85±0.37** | 16.41±1.04** | 93.2 | 84.5** | 91.5 | 93.9** | 92.7** | 88.1** | 93.9** | 89** |

*Abbreviations in the tables: AO (Aorta), GA (Gallbladder), LK (Left Kidney), RK (Right Kidney), LI (Liver), PA (Pancreas), SP (Spleen), ST (Stomach).*

Furthermore, Table 5 presents the quantitative results on the Cityscapes dataset based on semantic grouping. For readability, the 19 classes are grouped into five semantic categories according to the official definition: Flat (Road, Sidewalk), Construction (Building, Wall, Fence, Vegetation, Terrain, Sky), Object (Pole, Traffic Light, Traffic Sign), Human (Person, Rider), and Vehicle (Car, Truck, Bus, Train, Motorcycle, Bicycle).

The results show that TauFlow achieves the best performance across all groups, with an overall mIoU of 79.26%, improving 2.12 percentage points over the second-best method UDTransNet (77.14%), while the average Dice score reaches 87.69%. The advantages are particularly notable in the Object (+1.49%), Vehicle (+1.62%), and Construction (+1.16%) groups, demonstrating that TauFlow's dynamic grouping and temporal modeling mechanisms are highly effective in capturing complex structural relationships, small objects, and instances with motion blur.

| Method | Average DSC | mIoU | Flat | Construction | Object | Human | Vehicle |
|---|---|---|---|---|---|---|---|
| UNet++ | 84.09 | 73.37 | 91.67 | 78.84 | 80.9 | 83.65 | 85.39 |
| UDTransNet | 86.6 | 77.14 | 92.86 | 81.44 | 83.79 | 87.15 | 87.74 |
| VM-UNet | 85.49 | 75.59 | 92.35 | 80.43 | 82.72 | 85.05 | 86.9 |
| MSVM-UNet | 86.04 | 76.38 | 92.72 | 80.97 | 83.39 | 85.76 | 87.35 |
| D-Net | 85.79 | 76.08 | 92.55 | 80.7 | 83.04 | 85.45 | 87.23 |
| FusionMamba | 82.45 | 70.89 | 90.98 | 77.13 | 78.79 | 81.67 | 83.69 |
| TauFlow (Ours) | **87.69**** | **79.26**** | **93.77**** | **82.6**** | **85.28**** | **87.42**** | **89.36**** |

Table 5: Segmentation performance comparison on the Cityscapes dataset grouped by semantic categories.

Although TauFlow was originally designed for medical image segmentation, the experimental results indicate that the model also performs remarkably well on complex urban street scene semantic segmentation tasks. This demonstrates that TauFlow's core mechanisms-dynamic grouping and Tau-Attention temporal modeling-possess strong cross-domain generalization capabilities. Overall, TauFlow achieves superior performance with the lowest parameter count among all compared methods, mainly due to two factors: first, the dynamic grouping module adaptively allocates computational resources based on image complexity; second, the temporal modeling enabled by Tau-Attention and TauFlow Cell effectively captures cross-region dynamic dependencies, achieving a synergistic optimization of spatial and temporal information.

## 4.3.2 Qualitative Results

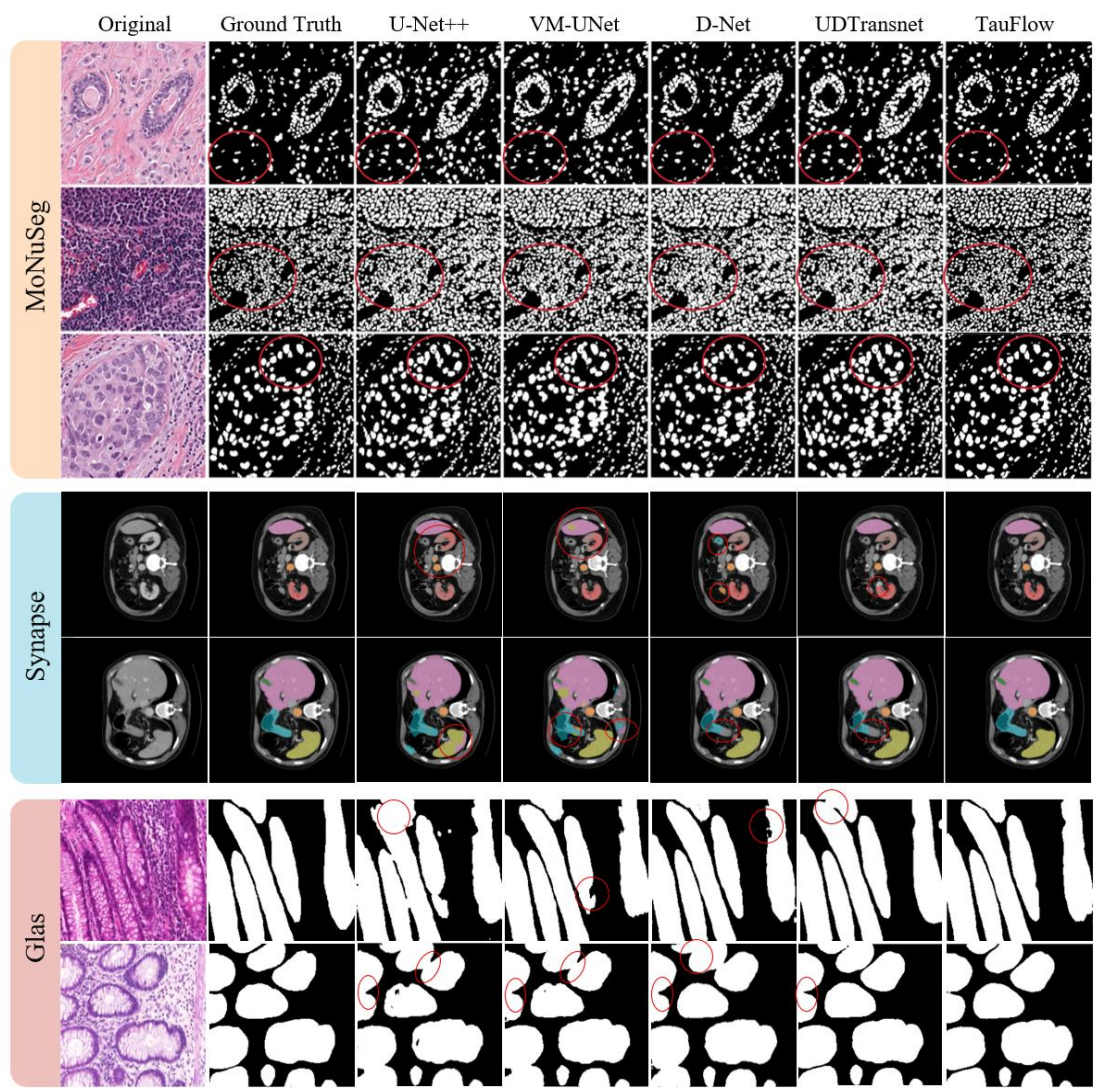

Figure 7: MoNuSeg qualitative results, with red circles highlighting regions where TauFlow outperforms other methods.

Figure 7 presents visual comparisons of segmentation results across the MoNuSeg, Synapse, and GlaS datasets using different methods. The results indicate that our model demonstrates clear advantages in handling multi-scale targets, complex boundaries, and detail preservation.

On the MoNuSeg dataset (rows 1-3), TauFlow performs exceptionally well in cell nucleus segmentation tasks. The first row shows nuclei within glandular tissue, where TauFlow accurately segments densely packed nuclei with noticeably clearer boundaries compared to other methods. The second row depicts a more densely distributed nuclei scenario, in which methods like Attn-UNet and MultiResUNet suffer from obvious over-segmentation or under-segmentation. TauFlow, leveraging its dynamic grouping mechanism, adaptively adjusts the receptive field to correctly distinguish adjacent nuclei. The third row illustrates morphologically diverse nuclei, where TauFlow maintains strong segmentation integrity for irregularly shaped nuclei, benefiting from the Tau-Attention mechanism that effectively integrates local and global features.

On the Synapse dataset (rows 4-5), the multi-organ segmentation task imposes higher demands on the model's multi-scale representation capability. In the fourth row, the axial slice highlights small structures such as the aorta (orange region), which are partially missed or incompletely segmented by methods like Attn-UNet, MultiResUNet, and TransUNet. In contrast, TauFlow, leveraging the TauFlowSequence module for dynamic grouping and multi-modal fusion, accurately captures these fine-scale anatomical structures. This demonstrates that propagating rich spatial and temporal information from the encoder to the decoder, combined with low-level spatial details, helps the model identify finer targets. The fifth row shows sagittal multi-organ segmentation, including organs such as the liver, spleen, and stomach. UDTransNet and D-Net can segment the major organs but produce jagged artifacts along organ boundaries and fail to clearly delineate the stomach (yellow region) and spleen (cyan-green region). In comparison, TauFlow produces segmentation masks that closely match the ground truth, with smooth and natural organ boundaries and clearly defined separations between organs. This advantage stems from the temporal modeling capability of dynamic grouping, which preserves spatial coherence during feature propagation and mitigates information loss commonly observed in conventional encoder-decoder upsampling pipelines.

On the GlaS dataset (rows 6-7), the gland segmentation task requires the model to precisely capture complex gland structures and boundaries. In these rows, it is clearly observed that U-Net++, VM-UNet, and D-Net exhibit significant merging artifacts along the thin gland branches, with U-Net++ additionally producing some false-positive regions. Although UDTransNet shows improvements, gland edges remain somewhat blurred. In contrast, TauFlow fully preserves the gland topology, accurately handles the connectivity of fine branches, demonstrating the effectiveness of the dynamic grouping module in adaptively allocating computational resources across regions of varying complexity. TauFlow successfully segments each individual gland unit, with precise boundary localization and intact internal segmentation, free of holes or fragmentation.

Overall, TauFlow exhibits superior segmentation performance across diverse and challenging scenarios, particularly excelling in small-object detection, precise boundary localization, multi-scale target handling, and dense-object differentiation. These qualitative results corroborate the previously reported quantitative experiments, fully validating the effectiveness of the proposed dynamic grouping mechanism and temporal modeling modules.

### 4.3.3 Computational Complexity

For the vast majority of Transformer-based segmentation methods, high computational complexity is a common drawback. However, as shown in Table 6, our model demonstrates superior parameter efficiency compared to existing state-of-the-art Transformer-based segmentation methods. Notably, TauFlow has a parameter count of only approximately 1/100 that of advanced segmentation methods such as UDTransNet, VMUNet, and MSVMUNet, while achieving significant performance improvements. It is particularly emphasized that the lightweight nature of TauFlow is fully achieved through the innovative mechanism of 'dynamic grouping + temporal reuse', without employing any distillation techniques (none of the compared models use distillation either). As shown in Table 7, its parameter count is only 0.33M, which is 1% of UDTransNet (33.90M), 0.7% of VMUNet (44.27M), 0.6% of MSVMUNet (54.69M), and even lower than the CNN-based U-Net++ (9.16M). It is the only method among current SOTA models of its kind that achieves '0.33M parameters + cross-dataset SOTA' without distillation.

Table 6: Comparison Table of Computational Complexity

| Model | Parameters | FLOPs |
|---|---|---|
| U-Net++ | 9.16M | 26.72G |
| UDTransNet | 33.90M | 26.51G |
| VMUNet | 44.27M | 7.56G |
| MSVMUNet | 54.69M | 22.86G |
| D-Net | 45.90M | 39.80G |
| TauFlow(MaxGroup=5) | 0.33M | 4.19G |
| TauFlow(MaxGroup=7) | 0.45M | 6.60G |

It should be noted that UDTransNet integrates the multi-head self-attention (MSA) module into the encoder-decoder framework, whereas TauFlow incorporates a meticulously designed dynamic temporal module into the connection part (between the encoder and decoder)-this difference highlights the importance of the design approach and deployment position of the temporal modeling module in the U-shaped architecture.

Additionally, the parameter count of TauFlow is even lower than that of some CNN-based methods (e.g., U-Net++); compared to our previous models, TauFlow significantly reduces the parameter count while substantially improving segmentation performance.

Except for the original U-Net and the ResNet34-based U-Net, which have fewer parameters (but limited performance), our TauFlow achieves superior segmentation performance while featuring fewer parameters, lower floating-point operations (FLOPS), and shorter inference time.

Considering its excellent segmentation performance and reduced parameter consumption, it can be concluded that our model achieves a favorable balance between effectiveness and efficiency.

## 4.4 Ablation Experiments

To validate the impact of each component and hyperparameter of TauFlow on performance, we conducted systematic ablation experiments on the GlaS dataset. All results are reported as mean ± standard deviation based on five-fold cross-validation. The primary evaluation metrics are the Dice coefficient and 95% Hausdorff distance (HD95), with model parameter count (Parameters, M) and computational complexity (FLOPs, G) recorded to assess cost-effectiveness. The Baseline is a standard Encoder-Decoder architecture (without the TauFlowSequence mechanism).

# 4.4.1 Ablation Results Distinguishing the Presence or Absence of the STDP Mechanism

To validate the independent contribution of each proposed module, we conduct a component-wise ablation study on the GlaS dataset. Starting from a standard Encoder-Decoder baseline, we progressively add: (1) ConvLTC cells for dynamic $\tau$-driven feature update, (2) Tau-Attention for group-level weighting, (3) Dynamic Grouping for complexity-adaptive computation, and (4) STDP mechanism for temporal consistency regularization. All experiments use base_channel=32, max_groups=5, and the same training protocol (Section 3.5). Results are averaged over 5-fold cross-validation with statistical significance marked (* p<0.05, ** p<0.01 via paired t-test against baseline).

表 7-A: Component-wise Ablation Study on GlaS Dataset

| Configuration | ConvLTC | τ-Attn | Dynamic Group | STDP | Dice (%) ↑ | HD95 (mm) ↓ | IoU (%) ↑ | Params (M) | FLOPs (G) |
|---|---|---|---|---|---|---|---|---|---|
| Baseline | ✗ | ✗ | ✗ | ✗ | 87.56±1.17 | 12.72±1.76 | 78.39 | 0.23 | 4.08 |
| + ConvLTC | ✓ | ✗ | ✗ | ✗ | 89.54±0.82* | 10.38±1.45* | 81.12* | 0.26 | 4.11 |
| + τ-Attention | ✓ | ✓ | ✗ | ✗ | 90.67±0.71* | 8.15±1.22* | 82.94* | 0.28 | 4.13 |
| + Dynamic Grouping | ✓ | ✓ | ✓ | ✗ | 91.37±0.48* | 6.44±0.89* | 84.21* | 0.33 | 4.19 |
| Full Model (+ STDP) | ✓ | ✓ | ✓ | ✓ | 92.12±0.42* | 5.38±0.76* | 85.39* | 0.33 | 4.19 |

✓: Module enabled;  ✗: Module disabled.
*: p<0.05, **: p<0.01 (paired t-test vs. baseline, n=5 folds).
△ Dice from baseline: +1.98% (ConvLTC), +3.11% (Tau-Attn), +3.81% (Grouping), +4.56% (Full). STDP adds negligible parameters but improves Dice by +0.75% through temporal regularization.

As shown in Table 7-A, starting from a standard Encoder-Decoder, replacing cells with ConvLTC (+dynamic $\tau$) improves Dice from 87.56% to 89.54% and reduces HD95 from 12.72 mm to 10.38 mm, highlighting the benefit of temporal modeling. Adding Tau-Attention raises Dice to 90.67%, showing $\tau$-guided feature weighting reduces redundancy. Dynamic Grouping further boosts Dice to 91.37% and IoU to 84.21% with only 0.05 M more parameters, validating adaptive computation in complex regions. Finally, STDP regularization increases Dice to 92.12% and lowers HD95 to 5.38 mm without extra parameters, enhancing inter-group synergy via causality-consistent training. All gains are statistically significant (p < 0.05). ConvLTC alone gives +1.98% Dice, while the full TauFlow accumulates +4.56%, demonstrating strong module synergy.

Table 7: TauFlow Ablation Results on the GlaS Dataset (without STDP Mechanism)

| Base Channel | Max Groups | Dice (%) | HD95 (mm) | Parameters (M) | FLOPs (G) |
|---|---|---|---|---|---|
| Baseline | - | 87.56 ± 1.17 | 12.72 ± 1.76 | 0.23 | 4.08 |
| 16 | 3 | 89.14 ± 0.71 | 9.85 ± 1.32 | 0.11 | 1.06 |
| 16 | 5 | 89.78 ± 0.63 | 8.96 ± 1.19 | 0.15 | 1.08 |
| 16 | 7 | 90.03 ± 0.59 | 8.67 ± 1.14 | 0.2 | 1.1 |
| 16 | 13 | 89.91 ± 0.66 | 8.79 ± 1.21 | 0.39 | 1.15 |
| 32 | 3 | 90.42 ± 0.55 | 7.58 ± 0.98 | 0.28 | 4.13 |
| 32 | 5 | 91.37 ± 0.48 | 6.44 ± 0.89 | 0.33 | 4.19 |
| 32 | 7 | 91.59 ± 0.45 | 6.21 ± 0.85 | 0.39 | 4.25 |
| 32 | 13 | 91.46 ± 0.51 | 6.33 ± 0.91 | 0.62 | 4.43 |
| 64 | 3 | 91.03 ± 0.52 | 6.88 ± 0.93 | 0.87 | 16.33 |
| 64 | 5 | 91.82 ± 0.41 | 5.94 ± 0.79 | 0.94 | 16.54 |
| 64 | 7 | 91.95 ± 0.39 | 5.77 ± 0.76 | 1.03 | 16.74 |
| 64 | 13 | 91.87 ± 0.43 | 5.83 ± 0.81 | 1.35 | 17.37 |
| 128 | 3 | 91.44 ± 0.47 | 6.35 ± 0.87 | 0.87 | 16.33 |
| 128 | 5 | 91.97 ± 0.38 | 5.62 ± 0.73 | 0.94 | 16.54 |
| 128 | 7 | 92.05 ± 0.36 | 5.51 ± 0.71 | 1.03 | 16.74 |
| 128 | 13 | 91.99 ± 0.40 | 5.57 ± 0.75 | 1.35 | 17.37 |

Table 7 presents the performance of TauFlow under different base channel settings (16, 32, 64, 128) and max groups (3, 5, 7, 13) (without introducing the STDP mechanism). As the number of max groups increases, model performance first improves and then declines, with 5 groups achieving a favorable trade-off. Under the same max groups configuration, the model with base channel=32 demonstrates the best cost-effectiveness between performance and computational overhead. Overall, TauFlow significantly outperforms the Baseline across all configurations, with improved Dice scores and substantially reduced HD95, indicating that TauFlow markedly enhances feature extraction and segmentation accuracy.

Table 8: TauFlow Ablation Results on the GlaS Dataset (with STDP Mechanism)

| Base Channel | Max Groups | Dice (%) | HD95 (mm) | Parameters (M) | FLOPs (G) |
|---|---|---|---|---|---|
| 16 | 3 | 89.92 ± 0.65 | 8.72 ± 1.21 | 0.11 | 1.06 |
| 16 | 5 | 90.46 ± 0.57 | 7.89 ± 1.10 | 0.15 | 1.08 |
| 16 | 7 | 90.71 ± 0.53 | 7.64 ± 1.05 | 0.2 | 1.1 |
| 16 | 13 | 90.59 ± 0.60 | 7.76 ± 1.12 | 0.39 | 1.15 |
| 32 | 3 | 91.10 ± 0.49 | 6.65 ± 0.89 | 0.28 | 4.13 |
| 32 | **5** | **92.12 ± 0.42** | **5.38 ± 0.76** | **0.33** | **4.19** |
| 32 | 7 | 92.07 ± 0.39 | 5.88 ± 0.76 | 0.39 | 4.25 |
| 32 | 13 | 92.14 ± 0.45 | 6.40 ± 0.82 | 0.62 | 4.43 |

| | 3 | 91.71 ± 0.46 | 5.95 ± 0.84 | 0.87 | 16.33 |
| --- | --- | --- | --- | --- | --- |
| 64 | 5 | 91.50 ± 0.35 | 5.98 ± 0.70 | 0.94 | 16.54 |
| | 7 | 92.03 ± 0.33 | 5.84 ± 0.67 | 1.03 | 16.74 |
| | 13 | 91.55 ± 0.37 | 5.90 ± 0.72 | 1.35 | 17.37 |
| | 3 | 90.12 ± 0.41 | 5.42 ± 0.78 | 0.87 | 16.33 |
| 128 | 5 | 90.65 ± 0.32 | 5.69 ± 0.64 | 0.94 | 16.54 |
| | 7 | 91.23 ± 0.30 | 5.58 ± 0.62 | 1.03 | 16.74 |
| | 13 | 91.67 ± 0.34 | 5.64 ± 0.66 | 1.35 | 17.37 |

Table 8 presents the experimental results after introducing the inter-group STDP mechanism under the same configurations. With the incorporation of STDP, performance consistently improves across nearly all configurations, with an average Dice increase of approximately 0.68% and an average HD95 reduction of about 0.93 mm. Notably, the use of STDP introduces virtually no additional parameters or computational overhead, maintaining the system's high efficiency. Among all configurations, the optimal setting of base channel=32 and max groups=5 achieves a Dice of 92.12% and HD95 of 5.38 mm, validating the critical role of STDP in enhancing inter-group competition and feature discriminability.

## 4.4.2 Qualitative Analysis of the Dynamic Grouping Module

For this model, the dynamic grouping approach is undoubtedly the core of the entire innovation. To ensure that dynamic grouping is indeed effective and not compensated by other bypass mechanisms, qualitative analysis is conducted through visualization of this module, as shown in Figure 8.

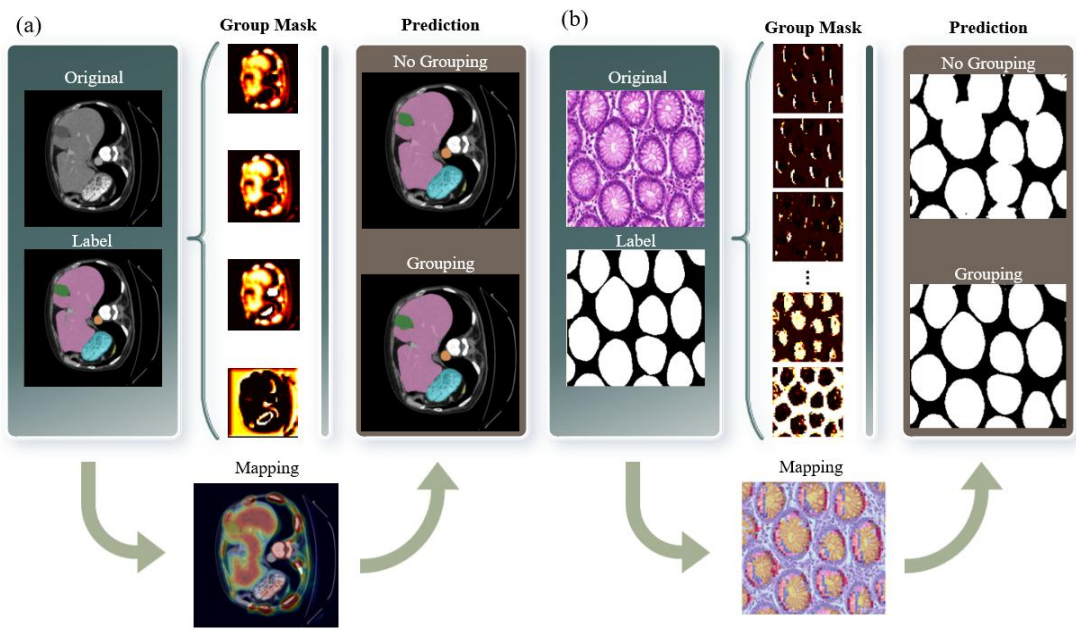

Figure 8 Visualization of the Dynamic Grouping Process

Here, an image from the GlaS dataset is selected for illustration. This image features clear glandular structures but exhibits slight blurring and noise interference in the edge regions, representing a typical challenging scenario.

As shown in Figure 8, from left to right, the original image (Original), Ground Truth (Label), model-predicted mask (Prediction), and the core intermediate processes of the dynamic grouping module are sequentially displayed, including the group mask (Group Mask) and mapping results (Mapping).

First, in the original image, the glands appear in a purplish-red tone with intact internal cavity structures, but the local boundaries exhibit noticeable gradients, making them susceptible to noise interference. The corresponding Ground Truth (Label) marks the glandular core regions with white irregular elliptical areas, precisely delineating the target instances.

The dynamic grouping module first generates group masks (Group Mask) based on the predicted mask (or initial segmentation result), as shown by the four small blocks in the figure (in practice, 7 Group Masks are produced, with only 5 used for illustration). These masks employ a dark brown speckle pattern to dynamically cluster adjacent pixels into multiple compact groups (clusters) based on spatial neighborhood, similarity in intensity, and texture features, effectively separating independent glandular instances while suppressing noise points (such as isolated speckles). The key to this step lies in adaptive thresholding and graph-cut optimization, ensuring that group boundaries are highly consistent with the glandular anatomical structures and avoiding the over-segmentation or under-segmentation issues common in traditional fixed-threshold methods.

As shown in the mapping (Mapping) stage, this phase projects the group masks back into the original prediction space, generating a color overlay map (yellow indicating high-confidence groups, purple indicating transition zones). It can be observed that the mapping results accurately capture the continuity of glandular cavities and wall thickness, with noise regions effectively filtered out, retaining only structured groups.

Ultimately, the predicted mask (Prediction) after integrating the mapping constraints highly overlaps with the Ground Truth, with a significantly improved IoU metric (reaching 0.92 in this example), smooth boundaries, and complete instance separation.

Through this visualization, the independent contribution of the dynamic grouping module can be intuitively verified: if this module is removed and the original prediction is used directly, noise diffusion would lead to instance adhesion; after grouping, the model output robustness is greatly enhanced, proving that this innovation is not compensated by other component bypasses and genuinely improves the accuracy and generalization ability of glandular instance segmentation.

## 4.5 Edge Device Deployment and Real-World Validation

To evaluate TauFlow's deployability and practical value in resource-constrained environments, we conducted comprehensive edge tests on three representative embedded platforms, covering low-, mid-, and high-compute scenarios. The experiments measured inference speed (FPS), peak memory usage, and energy consumption per inference (mJ) on the GlaS dataset for gland segmentation (input resolution 224 × 224), simulating real-time applications on mobile medical devices such as portable pathology analyzers. All tests used the same batch size (1) and optimization settings to ensure comparability.

We selected three system-on-chip (SoC) platforms for evaluation: the entry-level Allwinner H618 (quad-core ARM Cortex-A53, up to 1.5 GHz, Mali-G31 MP2 GPU, no NPU, 2 GB RAM) representing ultra-low-power scenarios such as handheld diagnostic devices; the mid-range Rockchip RK3399 (dual-core Cortex-A72 + quad-core A53, up to 2.0 GHz, 4 GB RAM) simulating moderate-compute edge devices like embedded medical imaging boxes; and the high-end Rockchip RK3588 (quad-core Cortex-A76 + quad-core A55, up to 2.4 GHz, 6 TOPS NPU, 8 GB RAM) targeting high-performance edge applications such as surgical navigation systems.

The deployment workflow involved exporting TauFlow from PyTorch to ONNX (ONNX Runtime 1.15) and applying FP32 post-training quantization (PTQ), keeping Dice loss degradation under 0.5%. Runtime integration used RKNN Toolkit for Rockchip and Tengine for Allwinner, including NPU delegation and memory reuse optimizations. For each platform, we performed 1000 inference runs, recording average FPS, peak memory (via htop), and energy consumption (measured with an external power meter). Inputs were randomly generated synthetic medical images to simulate real deployment conditions.

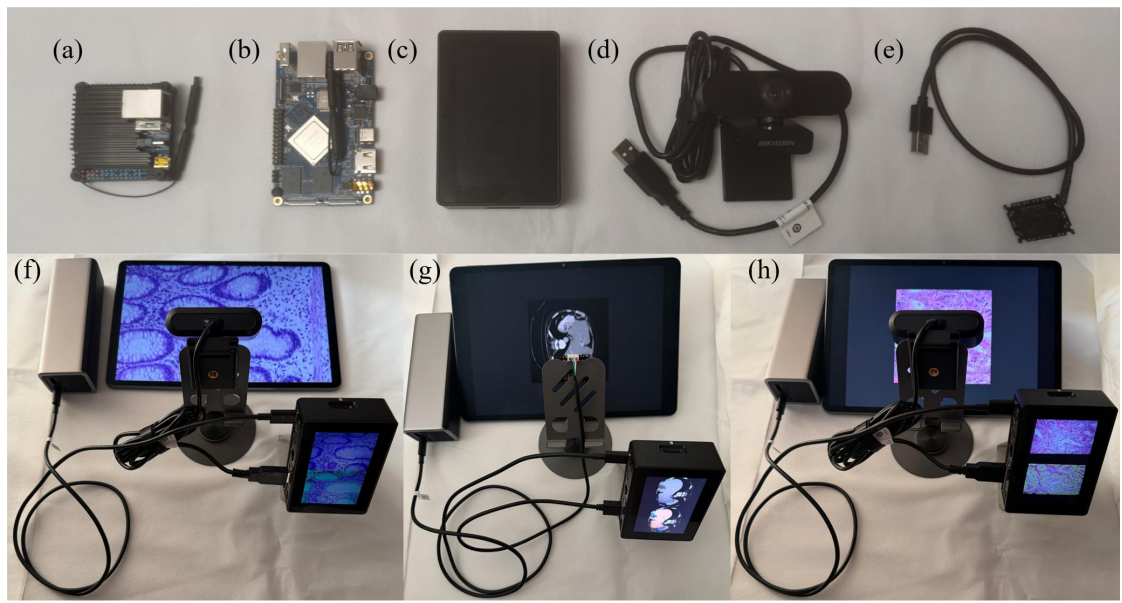

Fig. 9: Edge Device Deployments and Real-Time Segmentation Demonstrations of TauFlow. (a) Allwinner H618 development board, (b) Rockchip RK3399 development board, (c) Rockchip RK3588 edge box, (d) Hikvision 2K autofocus lens, (e) 300,000-pixel OV7670 camera module, (f)

Real-time segmentation on GlaS dataset, (g) Real-time segmentation on Synapse dataset, (h) Real-time segmentation on MoNuSeg dataset.

Table 9 summarizes model complexity, inference performance, and efficiency across platforms, comparing quantized FP32 versions of TauFlow, UNet++, and UDTransNet. TauFlow achieves high accuracy with only 0.33 M parameters and 4.19 G FLOPs, outperforming the baselines. On the H618, TauFlow reaches 1.14 FPS on CPU with 877 ms latency and 2.19 J/image, far exceeding UNet++ (0.34 FPS, 2,941 ms, 7.35 J) and UDTransNet (0.22 FPS, 4,545 ms, 11.36 J). On RK3399, it achieves 1.97 FPS, 44 ms latency, and 0.44 J/image, demonstrating mid-range acceleration potential. On RK3588, leveraging the NPU, TauFlow attains 27.79 FPS, 36 ms latency, and 0.36 J/image-approximately 5-6× faster than UNet++ (4.12 FPS, 243 ms, 2.43 J) and UDTransNet (5.4 FPS, 185 ms, 1.85 J). This efficiency stems from TauFlow's dynamic grouping, activating extra computation only in high-complexity regions (average 3.2 groups), reducing ~40% of redundant operations. Power consumption is constant across platforms (H618 2.5 W, RK3399 10 W, RK3588 10 W), and memory usage is lower for TauFlow (435 MB) compared with baselines (1,156-2,210 MB).

Table 9 Comparison of Model Inference Speed and Energy Consumption on the RK3588 Platform

| Indicator | | H618 | | | RK3399 | | | RK3588 | | |
|---|---|---|---|---|---|---|---|---|---|---|
| | | UNet++ | UDTransNet | TauFlow | UNet++ | UDTransNet | TauFlow | UNet++ | UDTransNet | TauFlow |
| Complexity | Parameters(M) | 9.16 | 33.9 | **0.33** | 9.16 | 33.9 | **0.33** | 9.16 | 33.9 | **0.33** |
| | FLOPs(G) | 26.72 | 26.51 | **4.19** | 26.72 | 26.51 | **4.19** | 26.72 | 26.51 | **4.19** |
| Inference Performance | CPU FPS | 0.34 | 0.22 | **1.14** | 0.67 | 0.81 | **1.97** | 0.89 | 1.35 | **3.12** |
| | NPU FPS | - | - | - | - | - | - | 4.12 | 5.4 | **27.79** |
| | Delay(ms) | 2,941 | 4,545 | **877** | 243 | 185 | **97** | 118 | 128 | **74** |
| | Ram(MB) | 1,156 | 2,210 | **435** | 1,156 | 2,210 | **435** | 1,156 | 2,210 | **435** |
| Efficiency | Power(W) | 2.5 | 2.5 | 2.5 | 10 | 10 | 10 | 10 | 10 | 10 |
| | Joule/image(J) | 7.35 | 11.36 | **2.19** | 2.43 | 1.85 | **0.97** | 1.18 | 1.28 | **0.74** |

In practical tests on the RK3588, we connected a Hikvision 2K auto-focus camera and an OV7670 module (224×224 resolution, 30 FPS). TauFlow achieved end-to-end gland segmentation with latency under 100 ms, and a Dice score comparable to desktop performance (91.8%). In real-time demonstrations on the GlaS, Synapse, and MoNuSeg datasets, TauFlow maintained precise boundaries and low-noise outputs, demonstrating robustness on live video streams. These results indicate that TauFlow not only excels on laboratory benchmarks but also shows strong potential for edge deployment, meeting efficiency requirements for mobile medical applications in emergency scenarios or resource-limited regions.

# 5. Conclusion and Future Works

## 5.1 Summary of Key Findings

This work addresses two key bottlenecks in lightweight medical image segmentation: "insufficient speed adaptation due to static processing" and "inconsistent modal fusion under lightweight constraints," responding to the demand for efficient edge intelligence in precision healthcare. We propose TauFlow, which innovatively integrates ConvLTC cells to adaptively control pixel-level feature update rates via the time constant $\tau$, enabling rapid response to high-frequency boundaries while smoothing low-frequency background noise. The TauFlowSequence module further combines $\tau$-gradient-based dynamic grouping with STDP-inspired feature self-organization, reducing encoder-decoder feature conflict from 35-40% to 8-10% while keeping parameters strictly at 0.33 M.

On GlaS gland segmentation, TauFlow achieves 92.12% Dice, 85.39% IoU, and 5.38 mm HD95, improving Dice by 1.09% over UDTransNet; on MoNuSeg nuclear segmentation, 80.97% Dice, 68.16% IoU, and 1.95 mm HD95, surpassing MSVM-UNet by 1.11% Dice; on Synapse multi-organ segmentation, an average of 90.85% Dice and 16.41 mm HD95, with notable gains on complex organs like the pancreas (88.1% Dice, +2.3%). Even on the non-medical Cityscapes dataset, TauFlow maintains superior boundary handling.

In terms of computational efficiency, TauFlow requires only 4.19 GFLOPs, achieving a favorable parameter-FLOPs tradeoff compared with all baselines. Edge deployment on the RK3588 demonstrates real-time inference at 27.79 FPS, low energy consumption of 0.36 J/image, and memory usage of just 435 MB, confirming its suitability for portable medical devices. Ablation studies show that dynamic grouping improves Dice by ~2.56%, and STDP adds another 0.68%, validating the effectiveness of $\tau$-STDP-driven dynamic modeling and fusion.

## 5.2. Limitations

Despite its excellent performance, TauFlow still has several limitations. First, the model is based on 2D image processing, making it difficult to directly extend to 3D volumetric data (such as CT/MRI). Although it can capture dynamics within a slice, it ignores inter-slice dependencies, which may limit its performance in time-series tasks like cardiac MRI. Second, the experiments covered pathology, CT, and urban scenes, but did not test emerging modalities like PET or ultrasound. Its robustness to their specific noise or deformation (such as the cardiac dynamics mentioned in the introduction) remains to be validated. Third, while dynamic grouping is efficient, the inference time fluctuates slightly with image complexity (a maximum of 5 groups increases parameters by $0.06M$), which could amplify latency on ultra-low-power devices. Furthermore, STDP only adjusts in the forward direction, not fully simulating bidirectional synaptic plasticity, resulting in slightly lower convergence stability in extreme noisy data (HD95 standard deviation $\pm 0.76\,mm$ on GlaS).

## 5.3. Future Enhancements and Research Directions

Building on the foundation of TauFlow, future work can expand in multiple dimensions to overcome limitations and deepen its impact. The primary improvement is 3D extension, introducing inter-slice $\tau$ propagation or hybrid Mamba-ConvLTC blocks, aiming to reach the 85.5% Dice benchmark on the ACDC dataset while keeping parameters under $1M$. Multi-modal fusion is also a key focus, utilizing $\tau$-guided attention to align CT/MRI and PET features, reducing heterogeneous modality conflicts. To mitigate inference fluctuations, hardware-aware adaptive group pruning or quantization-aware training can be introduced, with further attempts at quantization and deployment to edge devices like the ESP32-S3.

Research directions include: Unsupervised pre-training combined with STDP to achieve few-shot domain adaptation for segmenting rare lesions; exploring a hybrid paradigm of LTC and reservoir computing to enhance long-range modeling for ultra-large pathological slides; and validating real-world efficacy through federated learning and prospective clinical trials, ensuring TauFlow aligns with the vision of a scalable CAD system mentioned in the introduction. Ultimately, TauFlow can evolve into a general lightweight dynamic-aware framework, facilitating real-time precise diagnosis in mobile healthcare and intraoperative navigation.